\pdfoutput=1

\documentclass[iop, apj]{emulateapj}

\usepackage{xspace}
\usepackage{amsmath}
\usepackage{framed} 
\usepackage{txfonts}
\usepackage{epstopdf} 
\usepackage{color}
\usepackage{rotating}
\usepackage{natbib}
\usepackage{ulem}
\usepackage{xspace}
\usepackage[usenames,dvipsnames,svgnames,table]{xcolor}

\slugcomment{Accepted for publication in the Astrophysical Journal}

\newdimen\hssize
\newcommand{\mpc}{\>{\rm Mpc}}

\newcommand{\kpch}{\>{h^{-1}{\rm kpc}}}
\newcommand{\mpch}{\>h^{-1}{\rm {Mpc}}}

\newcommand{\msun}{\>{\rm M_{\odot}}}

\newcommand{\kmsmpc}{\>{\rm km}\,{\rm s}^{-1}\,{\rm Mpc}^{-1}}

\def\gcm3{\mathrm{g} / \mathrm{cm}^3}

\def\gtsima{$\; \buildrel > \over \sim \;$}
\def\ltsima{$\; \buildrel < \over \sim \;$}
\def\prosima{$\; \buildrel \propto \over \sim \;$}
\def\gsim{\lower.7ex\hbox{\gtsima}}
\def\lsim{\lower.7ex\hbox{\ltsima}}
\def\simgt{\lower.7ex\hbox{\gtsima}}
\def\simlt{\lower.7ex\hbox{\ltsima}}
\def\simpr{\lower.7ex\hbox{\prosima}}

\def\rmd{{\rm d}}

\def\rml{{\rm l}}
\def\rmm{{\rm m}}

\def\rmp{{\rm p}}

\def\rmr{{\rm r}}
\def\rms{{\rm s}}

\shorttitle{
Weak Lensing and Clustering Measurements of BOSS Galaxies
}
\shortauthors{Miyatake~et~al.}

\begin{document}


\def\figdir{.}
\def\figext{eps}


\title{ The weak lensing signal and the clustering of BOSS galaxies I:
Measurements }

\author{Hironao~Miyatake \altaffilmark{1,2}, Surhud~More
\altaffilmark{2}, Rachel~Mandelbaum \altaffilmark{3}, Masahiro~Takada
\altaffilmark{2}, David~N.~Spergel \altaffilmark{1}, Jean-Paul~Kneib
\altaffilmark{4,5}, Donald~P.~Schneider \altaffilmark{6,7}, J.~Brinkmann
\altaffilmark{8}, Joel~R.~Brownstein \altaffilmark{9}} \affil{ $^1$
Department of Astrophysical Sciences, Princeton University, Peyton Hall,
Princeton NJ 08544, USA\\ $^2$ Kavli Institute for the Physics and
Mathematics of the Universe (WPI), The University of Tokyo, Chiba
277-8582, Japan\\ $^3$ McWilliams Center for Cosmology, Department of
Physics, Carnegie Mellon University, Pittsburgh, PA 15213, USA\\ $^4$
Aix Marseille Universit\'{e}, CNRS, LAM (Laboratoire d'Astrophysique de
Marseille) UMR 7326, 13388, Marseille, France\\ $^5$ Laboratoire
d'astrophysique, \'{E}cole Polytechnique F\'{e}d\'{e}rale de Lausanne
(EPFL), Observatoire de Sauverny, 1290 Versoix, Switzerland\\ $^6$
Department of Astronomy and Astrophysics, The Pennsylvania State
University, University Park, PA 16802, USA\\ $^7$ Institute for
Gravitation and the Cosmos, The Pennsylvania State University,
University Park, PA 16802, USA\\ $^8$ Apache Point Observatory, PO Box
59 Sunspot, NM 88349, USA\\ $^9$ Department of Physics and Astronomy,
University of Utah, 115 S 1400 E, Salt Lake City, UT 84112, USA }


\begin{abstract}

 A joint analysis of the clustering of galaxies and their weak
 gravitational lensing signal is well-suited to simultaneously constrain
the galaxy-halo connection as well as the cosmological parameters by
 breaking the degeneracy between galaxy bias and the amplitude of
 clustering signal. In a series of two papers, we perform such an
 analysis at the highest redshift ($z\sim0.53$) in the literature using
 CMASS galaxies in the Sloan Digital Sky Survey-III Baryon Oscillation
 Spectroscopic Survey Eleventh Data Release (SDSS-III/BOSS DR11) catalog
 spanning 8300~deg$^2$. In this paper, we present details of the
 clustering and weak lensing measurements of these galaxies. We define a
 subsample of 400,916 CMASS galaxies based on their redshifts and
 stellar mass estimates so that the galaxies constitute an approximately
 volume-limited and similar population over the redshift range $0.47\le
 z\le 0.59$. We obtain a signal-to-noise ratio $S/N\simeq 56$ for the
 galaxy clustering measurement. We also explore the redshift and stellar
 mass dependence of the clustering signal. For the weak lensing
 measurement, we use existing deeper imaging data from the CFHTLS with
 publicly available shape and photometric redshift catalogs from
 CFHTLenS, but only in a 105~deg$^2$ area which overlaps with BOSS. This
 restricts the lensing measurement to only 5,084 CMASS galaxies. After
 careful systematic tests, we find a highly significant detection of the
 CMASS weak lensing signal, with total $S/N\simeq 26$. These
 measurements form the basis of the halo occupation distribution and
 cosmology analysis presented in More et al. (Paper II).

\end{abstract}


\section{Introduction}

In the current concordance cosmological model ($\Lambda$CDM), dark
matter and dark energy constitute a large fraction ($\sim95.5\%$) of the
energy density of the Universe; this conclusion is supported by a
growing body of diverse observational astrophysical evidence \citep[see
e.g.,][]{Hinshaw:2013, Planck:2013, Sullivan:2011, Suzuki:2012,
Rest:2013, Blake:2011, Beutler:2011, Anderson:2012, Aubourg:2014,
Reidetal:12}. Yet, we have little theoretical understanding of the
fundamental physics that governs the physics of dark matter or dark
energy. For this reason a large number of wide-area galaxy surveys
are ongoing or planned for the near future aimed at characterizing the
dark matter distribution and understanding the nature of dark energy.
These include both the imaging and spectroscopic surveys: the
Kilo-Degrees Survey
(KIDS)\footnote{\url{http://www.astro-wise.org/projects/KIDS/}}, the
Subaru Hyper Suprime-Cam (HSC)
Survey\footnote{\url{http://www.naoj.org/Projects/HSC/index.html}}
\citep[see also][]{Miyazakietal:12}, the Dark Energy Survey
(DES)\footnote{\url{http://www.darkenergysurvey.org}}, Extended Baryon
Oscillation Spectroscopic Survey
(eBOSS)\footnote{\url{https://www.sdss3.org/future/eboss.php}}, Subaru
Prime Focus Spectrograph (PFS)
Survey\footnote{\url{http://sumire.ipmu.jp/pfs/intro.html}} \citep[see
also][]{Takadaetal:14}, Dark Energy Spectroscopic Instrument
(DESI)\footnote{\url{http://desi.lbl.gov}}, and ultimately the Large
Synoptic Survey Telescope
(LSST)\footnote{\url{http://www.lsst.org/lsst/}} \citep[see
also][]{LSST_Science_Collaboration:2009}, the Euclid
project\footnote{\url{http://sci.esa.int/science-e/www/area/index.cfm?fareaid=102}},
and the Wide-Field Infrared Survey Telescope (WFIRST)
project\footnote{\url{http://wfirst.gsfc.nasa.gov}} \citep[see
also][]{WFIRST}.

The measurement of galaxy clustering statistics is one of the most
powerful probes in observational cosmology
\citep{Tegmarketal:04,Coleetal:05,Percivaletal:10,Zehavietal:11,Saitoetal:11,Reidetal:12,Reidetal:14,Samushia:2014}. However,
our lack of a detailed understanding of the relationship between the
distribution of galaxies and that of dark matter limits the full use of
the measured amplitude of the clustering signal for constraining
cosmological parameters. \citet{Seljak:2005} first proposed and
demonstrated that it is possible to break this degeneracy between the
unknown bias between galaxies and matter and the cosmological parameters
by utilizing the theoretical dependence of galaxy bias on halo mass
\citep{MoWhite:96,ShethTormen:99}. By probing the halo masses of
galaxies via the weak gravitational lensing around galaxies,
commonly referred to as galaxy-galaxy lensing, on small scales
\citep[see also][]{Yoo:2006, Cacciato:2009, Li:2009, Tinker:2012,
Leauthaud:2012, More:2013,Cacciato:2013b, Gillis:2013, Cacciato:2013,
Simpson:2013} \citep[see also][for the galaxy-galaxy lensing
measurements]{Fischer:2000,
Mandelbaum:2006,van_Uitert:2011,Velander:2013}.  An alternative method,
which uses the ratio of the clustering and lensing signals, has
also been proposed in order to avoid the complex astrophysics that
complicates the interpretation of these observables on scales smaller
than the typical halo radii \citep{Baldauf:2010}. As a proof of this
concept, \citet[][hereafter RM13]{Mandelbaum:2013} used the
state-of-the-art measurement of galaxy-galaxy lensing signal and galaxy
clustering up to large scales ($\sim70\mpch$) by combining the
spectroscopic and multi-color imaging galaxy catalogs from the Sloan
Digital Sky Survey I/II \citep[SDSS I/II; ][]{York:2000,
Eisenstein:2001, Strauss:2002}. They showed that the joint analysis
provides a significant improvement in the cosmological parameters
$\Omega_\rmm$ and $\sigma_8$ when combined with results from the
WMAP7 experiment \citep{Komatsu:2011}.

The purpose of this paper and our companion paper \citep[][hereafter
Paper II]{Moreetal:14} is to extend the joint clustering and lensing
analysis to galaxies at $z\sim 0.5$ and then place constraints on cosmological
parameters $\Omega_m$ and $\sigma_8(z)$ at the redshift. This study is at the
highest redshift among such kind of measurements in the literature. To do
this, we use independent data sets to measure the clustering and weak lensing
signals: the Baryon Oscillation Spectroscopic Survey
\citep[BOSS;][]{Schlegel:2010,Dawson:2013}, which is in a part of the SDSS-III
\citep[SDSS-III;][]{Eisenstein:2011}, and  the publicly-available
Canada-France-Hawaii Telescope Lensing Survey (CFHTLenS; \citealt{Heymans:2012})
catalog.

The BOSS survey is measuring spectroscopic redshifts of 1.5
millions galaxies, approximately volume limited to $z\simeq 0.6$. The
BOSS galaxies, the so-called CMASS (``constant mass'') galaxies, are
selected from the multi-color SDSS imaging data based on the
requirements that the number density of the CMASS galaxies, with
successful redshifts measurements, is high enough to probe large-scale
structure at an intermediate redshift around $z\sim 0.5$. The CFHTLenS
catalog contains galaxies with a median redshift of $z\sim 0.7$, where
every galaxy has its shape and photometric redshift (photo-$z$)
information estimated using the carefully designed
point-spread-function-matched photometry of different passband data
\citep{Hildebrandt:2012}. Thus the CFHTLenS catalog is best suited for
our purpose because it is much deeper and therefore contains a higher
number density of source galaxies than in the SDSS
\citep{Abazajian:2009}. In addition the photo-$z$ information
allows a selection of source galaxies which can be used to measure the
weak gravitational lensing signal of the CMASS galaxies, even though the
overlapping region of the two surveys is only $\sim$105 square degrees,
about one hundredth of the SDSS area.

In our analysis, we define subsamples of the parent CMASS
catalog based on their stellar mass estimates so that the galaxies
constitute a physically similar population over the redshift range
$0.47\le z\le 0.59$.  We will use the different subsamples, defined with
different stellar mass thresholds and in different redshift bins, to
test how the clustering and lensing signals of galaxies vary as a
function of redshift and stellar mass. Then we will use different stellar mass
threshold subsamples to test possible effects of the incompleteness or selection
inhomogeneities at the low stellar mass end on our cosmological
constraints. This paper, the first in a series of two, will present the details
of the subsamples and the clustering and lensing measurements. In Paper II we
will present the constraints on the halo occupation distribution of galaxies and
cosmological parameters derived from the measurements.

The structure of this paper is as follows. In
Section~\ref{sec:boss_data_and_clustering_measurement}, we describe the
BOSS data used in this paper, the definition of our subsamples
constructed from the parent CMASS catalog,
and details of the clustering signal measurements.
In Section~\ref{sec:cfhtlens_and_lensing_measurement}, we
describe the CFHTLenS data used in our lensing
measurement, details of our weak lensing analysis methodology, and the systematic
tests of the CFHTLenS catalog using random catalogs.
Section~\ref{sec:conclusion} is devoted to discussion and conclusions.
Unless stated otherwise, we will adopt a flat $\Lambda$CDM cosmology
with $\Omega_m=0.27$, $\Omega_\Lambda=0.73$, and the Hubble parameter
$h=H_0/(100\kmsmpc)=0.703$.

\section{SDSS-III BOSS data and clustering measurements}
\label{sec:boss_data_and_clustering_measurement}

\subsection{SDSS-III BOSS galaxies}
\label{sec:BOSS_data}

For the clustering measurements, we use the sample of galaxies
compiled in Data Release 11 (DR11) of the SDSS-III project. The SDSS-III
is a spectroscopic investigation of galaxies and quasars selected from
the imaging data obtained by the SDSS \citep{York:2000} I/II covering
about $11,000$~deg$^2$ \citep{Abazajian:2009} using the dedicated 2.5-m
SDSS Telescope \citep{Gunn:2006}. The imaging employed a drift-scan
mosaic CCD camera \citep{Gunn:1998} with five photometric bands ($u, g,
r, i$ and $z$) \citep{Fukugita:1996, Smith:2002, Doi:2010}. The SDSS-III
\citep{Eisenstein:2011} BOSS project \citep{Ahn:2012, Dawson:2013}
obtained additional imaging data of about 3,000~deg$^2$
\citep{Aihara:2011}. The imaging data was processed by a series of
pipelines \citep{Lupton:2001, Pier:2003, Padmanabhan:2008} and corrected
for Galactic extinction \citep{Schlegel:1998} to obtain a reliable
photometric catalog. This catalog was used as an input to select targets
for spectroscopy \citep{Dawson:2013} for conducting the BOSS survey
\citep{Ahn:2012} with the SDSS spectrographs \citep{Smee:2013}. Targets
are assigned to tiles of diameter $3^\circ$ using an adaptive tiling
algorithm designed to maximize the number of targets that can be
successfully observed \citep{Blanton:2003}. The resulting data were
processed by an automated pipeline which performs spectral
classification, redshift determination, and various parameter
measurements, e.g., the stellar mass measurements from a number of
different stellar population synthesis codes which utilize the
photometry and redshifts of the individual galaxies
\citep{Bolton:2012}. In addition to the galaxies targetted by the BOSS
project, we also use galaxies which pass the target selection but have
already been observed as part of the SDSS-I/II project (legacy
galaxies).  These legacy galaxies are subsampled in each sector so that
they obey the same completeness as that of the CMASS sample
\citep{Anderson:2014}.

\begin{figure*} \centering{
 \includegraphics[]{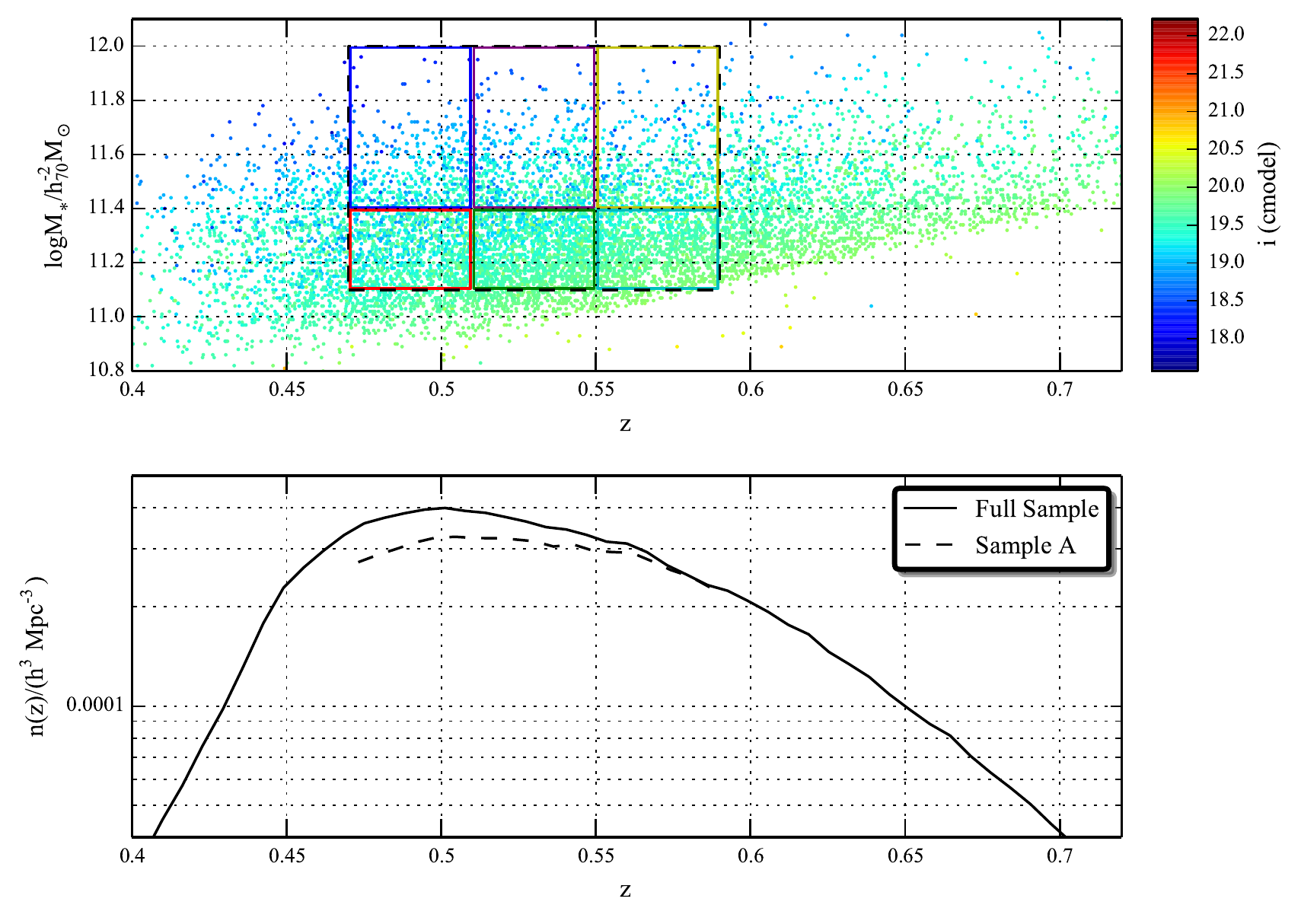}}
 \caption{ {\it Upper panel:} The CMASS galaxy sample in the
 redshift--stellar mass plane. The fiducial subsample (here after
 subsample A) used in this study corresponds to the redshift range
 $z\in [0.47,0.59]$ and stellar mass range $\log M_*\in[11.1,12.0]$,
 which is shown as the black bashed box. To investigate the redshift and
 stellar mass dependence of the clustering we further subdivide this
 subsample into six bins indicated by different colour boxes in
 this figure. In addition, we will investigate the robustness of the
 cosmological constraints with different subsamples that span the
 same redshift range as our fiducial subsample but progressively
 restricting the samples to more massive galaxies, $\log
 M_*\in[11.3,12.0]$ and $\log M_*\in[11.4,12.0]$. Hereafter we call
 these subsamples of higher stellar mass thresholds, subsample B and C,
 respectively.  To avoid crowding we show only 10,000 galaxies
 from the total CAMASS galaxies in this figure. Colors of each
 point denote the $i-$band cmodel magnitudes in the SDSS imaging
 catalog. {\it Lower panel:} The redshift dependence of the comoving
 number density. The dashed curve shows our fiducial subsample, which is
 almost constant in the redshift range centered at $z\simeq0.53$. For
 comparison, the solid curve shows the full CMASS sample.}
 \label{fig:Mstel_z}
\end{figure*}

To perform measurements of the clustering and lensing signals, we create
various subsamples of the parent large scale structure catalog provided
with DR11. To define the subsamples we use for the analysis, we make use
of the stellar masses processed through the Portsmouth stellar
population synthesis code \citep{Maraston:2013} with the assumptions of
a passively evolving stellar population synthesis model and a
\citet{Kroupa:2001} initial mass function.  The upper panel of
Figure~\ref{fig:Mstel_z} shows the distribution of a random subsample of
all galaxies in the large scale structure catalog in the stellar
mass-redshift plane. The number density of this sample varies as a
function of redshift and peaks at $z\sim0.5$, as shown in the
bottom panel. The fiducial subsample we use in this paper, denoted as
subsample A, consists of all galaxies between $z\in[0.47,0.59]$ and with
$\log M_\ast/h^{-2}_{70}\msun\in[11.1,12.0]$. This selection results in
a sample with an approximately uniform number density with redshift
compared to the parent sample, as shown by the black dashed line in
the lower panel of Figure~\ref{fig:Mstel_z}. The number density of this
fiducial subsample is $\sim 3 \times 10^{-4}~h^{3} \mpc^{-3}$. The total
number of galaxies in this subsample is $378~807$
$(400~916)$\footnote{The number in parentheses includes galaxies with
redshift failures and those that could not be allocated fibers to
measure their redshifts (fiber collided galaxies) as described in the
next paragraph.} and constitutes about half of the parent sample used
for the measurements of the baryon acoustic oscillations. In order to
test the effects of incompleteness at the low stellar mass end in our
fiducial subsample, we additionally consider two different subsamples of
galaxies which lie in the same redshift range but where the stellar mass
selection is $\log M_*/h^{-2}_{70}\msun\in [11.30,12.0]$ and $\log
M_*/h^{-2}_{70}\msun\in [11.40,12.0]$, respectively. These subsamples
will be denoted as $B$ and $C$, respectively. The numbers of galaxies in
these subsamples are $196~578$ and $116~682$ (these numbers include
fiber collided and redshift failure galaxies), while the number
densities of the galaxies are $1.5\times10^{-4}$ and $0.8\times10^{-4}
h^{3} \mpc^{-3}$, respectively. The number density of galaxies is
approximately constant in the redshift range we use. In addition to
these subsamples, we will also consider subsamples in different redshift
bins to test the redshift and stellar mass dependence of the clustering
signals.

We must account for a number of subtle selection effects in order to
obtain a precise measurement of clustering \citep{Ross:2012}. The
spectroscopic target sample is obtained from the SDSS imaging
observations after the application of a variety of colour and
photometric selection cuts \citep{Dawson:2013}.  However, due to the
limited number of fibers available, not all galaxies from this target
sample can be allocated a fiber while performing spectroscopic
observations to determine their redshifts. This could also happen if two
targets are within $62''$ of each other and hence they cannot be
simultaneously observed due to the finite size of fibers. If such
fiber-collided galaxies lie in a region of the sky which is visited
multiple times (due to overlaps in the target tiling) then they may have
redshift measurements.  There are also instances where a galaxy is
assigned to a fiber, but its redshift could not be obtained. Finally,
there are also instances where it is difficult to perform star-galaxy
separation, especially in fields with a high number density of stars.
These effects have been quantified in the parent DR11 catalog of CMASS
galaxies by assigning a weight to each galaxy such that
\begin{equation}
w_\rml = w_\ast \, ( w_{\rm noz} + w_{\rm cp} - 1 ) \,,
\end{equation}
where $w_{\rm noz}$ is the weight assigned to a galaxy if it is the
nearest neighbour (in the plane of the sky) of a redshift failure
galaxy, $w_{\rm cp}$ is similarly assigned to account for the nearest
neighbours of fiber collided galaxies\footnote{Nearest neighbour corrections
have been shown to accurately correct for fiber collisions above the
fiber collision scale ($\sim 0.4 \mpch$) by \citet{Guo:2012}. Both $w_{\rm noz}$ and
$w_{\rm cp}$ are equal to unity by default for all galaxies. Their
values are incremented for the nearest neighbours of every redshift
failure or fiber collided galaxy.}, and $w_\ast$ accounts for the
systematic relationship between density of stars and density of BOSS
target galaxies \citep[for details, see][]{Anderson:2014}. The BOSS
parent catalog contains an additional weight, $w_{\rm FKP}$, for each galaxy
which depends upon the number density of galaxies in the sample at its redshift
\citep{Feldman:1994}. This weight is important for the parent galaxy catalog
which has a much larger variation in the number density of galaxies than the
variation in $\bar{n}(z)$ for the subsamples of galaxies we use. Therefore, in
our analysis, we do not include this weight.

The weights $w_{\rm noz}$ and $w_{\rm
cp}$ can only be used if the entire sample of galaxies within a given
redshift range is used to measure the clustering signal. In
particular, for the subsamples of galaxies selected by stellar mass
(or luminosity), it is unclear whether the fiber-collided or
redshift-failure galaxy will be part of our subsample if it is
assigned the same redshift as its nearest neighbour, as it may or may
not satisfy the stellar mass cut we have imposed. If we were to use
the weights $w_{\rm noz}$ and $w_{\rm cp}$ as provided, then we would
spuriously include the weights of some galaxies which should
not be in the subsample. We will also miss some fiber-collided or
redshift-failure galaxies which should have been part of our subsample because
their nearest neighbours failed to make it to our subsample due to stellar mass
cuts.  In addition there is a possibility that the small scale clustering will
be affected by using the weights, as all pairs of galaxies involving a
fiber-collided or redshift-failure galaxy are assigned line-of-sight separations
equal to zero.

Given these issues, we refrain from using the weights $w_{\rm noz}$ and
$w_{\rm cp}$ in our analysis. Instead, we have obtained the stellar masses
using the measured photometry of the fiber-collided and the redshift-failure
galaxies with the redshift of its nearest neighbour.  Each of these galaxies are
assigned the same $w_\ast$ as their nearest neighbours. 
We have verified that
the stellar mass-redshift distribution of such galaxies is similar to that
of the sample of galaxies which have well measured redshifts (catastrophic
failures in the nearest neighbour redshift assumption will result in both a
an incorrect redshift and stellar mass). We then decide whether to include
these galaxies in our subsample based on whether these galaxies pass the
stellar mass cuts we impose.
Given our treatment of fiber collided galaxies, the only weight we have to
use in our analysis is 
\begin{equation}
\label{eq:lens_weight_cmass}
w_\rml = w_\ast.
\end{equation}

\subsection{CMASS galaxy clustering measurements}
\label{sec:clustering}

The clustering of galaxies can be quantified using the two-point correlation
function. The two-point correlation function, $\xi(r)$, depends only upon the
true three-dimensional distance between galaxies, $r$, if the Universe is
isotropic. However, the assumption of isotropy is broken due to
the modulation of the distances of galaxies along the line-of-sight ($\pi$)
caused by the peculiar motions of galaxies. In contrast, the distances along the
plane of the sky ($r_\rmp$) do not suffer from this modulation. The resultant correlation function displays a
characteristic anisotropic pattern which elongates (flattens) the
iso-correlation contours in the ($r_\rmp, \pi$) plane on small (large)
projected scales. The impact of such effects can be minimized by focusing on
the projected two-point correlation function obtained by integrating the
correlation function $\xi(r_\rmp,\pi)$ along the line-of-sight,
\begin{equation}
w_\rmp(r_\rmp)=2 \int_{0}^{\pi_{\rm max}} \xi(r_\rmp,\pi) \,\rmd \pi\,.
\label{eq:wp}
\end{equation}
Unless stated otherwise, we will use $\pi_{\rm max}=100 \mpch$, and employ
the same finite line-of-sight integration limit while analytically modelling the
observations in Paper II. When calculating the integral, we adopted
the binning of $\Delta\pi=1 \mpch$.

We use the estimator proposed by \citet{Landy:1993},
\begin{equation}
\xi(r_\rmp,\pi)=\frac{DD-2DR+RR}{RR}\,
\end{equation}
to obtain the two-point correlation function $\xi(r_\rmp,\pi)$. Here, $DD$, $RR$
and $DR$ represent the number of appropriately weighted pairs of galaxies with a
given separation $(r_\rmp,\pi)$, where both galaxies lie
either in the galaxy catalog or the random catalog or one in each of the
catalogs, respectively.

We use random catalogues with the same angular and redshift selection as
our galaxy subsample. These random catalogs consist of about 50
times more points than the number of galaxies in each of our
subsamples. We assign each random point a weight of $N_{\rm
gal}/N_{\rm ran}$ to account for this difference. In practice, we use
the random catalogs provided with SDSS DR11
\citep{Anderson:2014}. We subsample these random catalogs in order
to tailor them to the CMASS subsamples we use. For this purpose, we
first divide the entire CMASS catalog into narrow redshift bins ($\Delta
z=0.05$). For each bin we calculate the fraction of galaxies in our
subsample after the stellar mass and redshift cuts are applied to the
total sample, and interpolate this fraction as a function of
redshift. This fraction is used at each redshift as the probability to
accept points from the random catalog.  This procedure also
automatically accounts for the redshift cuts as the fraction outside the
redshift range we have chosen is identically equal to zero.

\begin{figure*} \centering{
\includegraphics[]{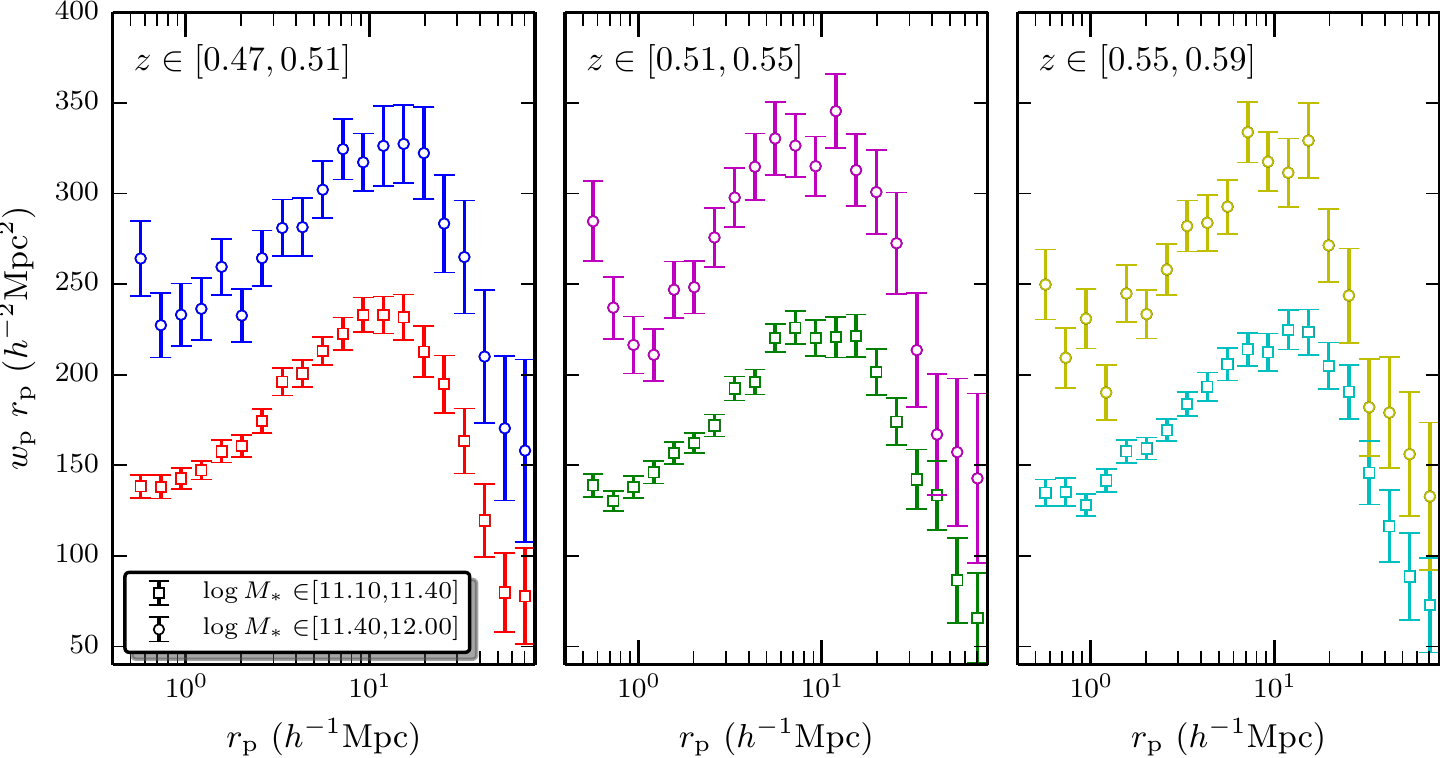}}
\caption{
The stellar mass dependence of the projected clustering signal at fixed
redshift is displayed in each of the panels for three different redshift bins.
Data points with errorbars show the clustering measurements of the different
subsamples shown in Figure~\ref{fig:Mstel_z}. Squares with errorbars correspond
to the stellar mass bin $\log M_*/h_{70}^{-2}M_\odot\in[11.10,11.40]$ while
circles with errorbars correspond to the stellar mass bin $\log
M_*/h_{70}^{-2}M_\odot\in[11.40,12.00]$. The line-of-sight integration length
used to project the redshift space correlation function is $60 \mpch$. At fixed
redshift, the clustering amplitude increases strongly with the stellar mass of
galaxies.
}
\label{fig:wp_fix_z}
\end{figure*}

\begin{figure*} \centering{
\includegraphics[]{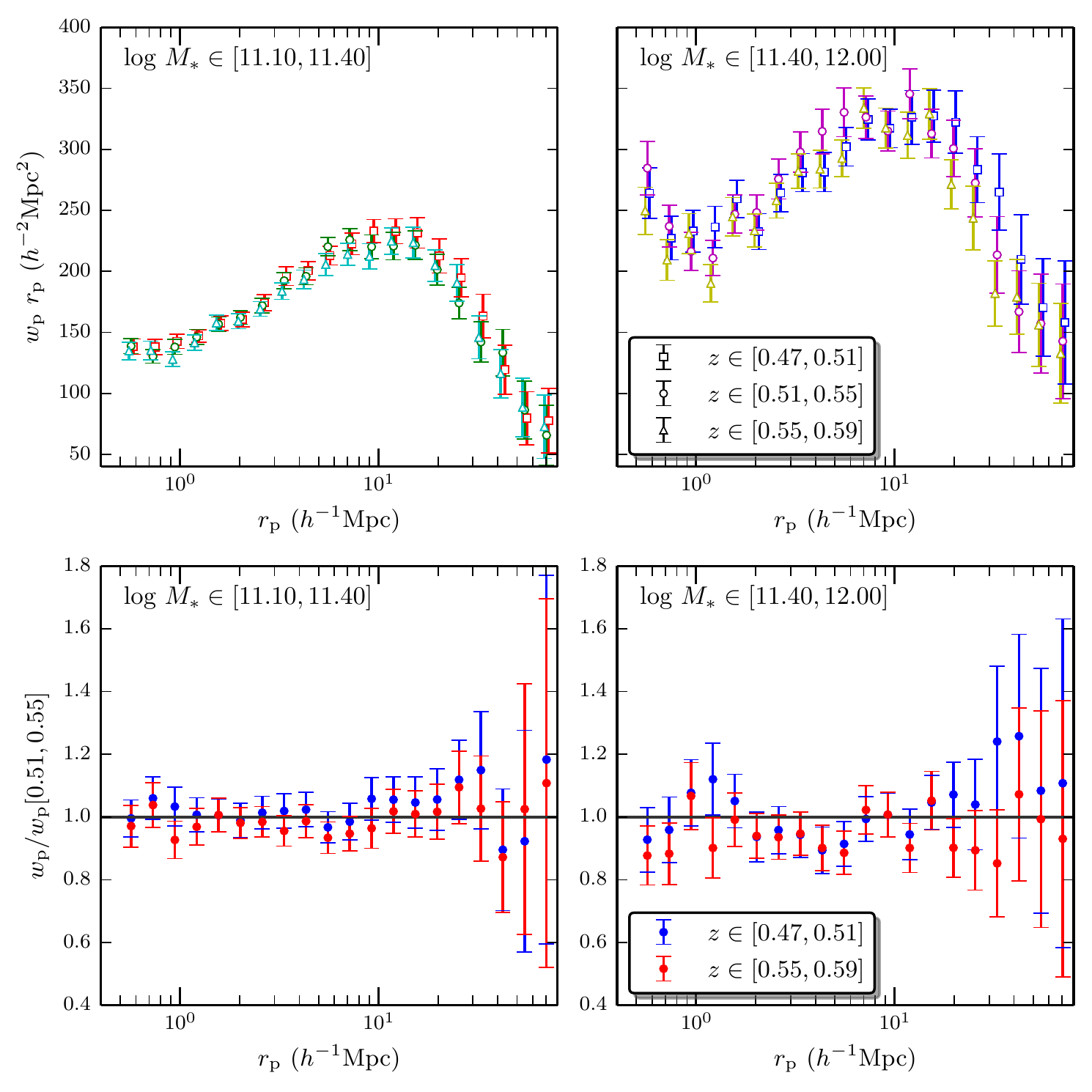}}
\caption{The upper left and right panels display the redshift dependence of the
clustering signal at fixed stellar mass for two different stellar mass bins.
Data points with errorbars show the clustering measurements of the different
subsamples shown in Figure~\ref{fig:Mstel_z}. In each panel, squares with
errorbars, circles with errorbars, and triangles with errorbars are used to
denote the measurement of clustering in the redshift bins
$[0.47,0.51],[0.51,0.55]$ and $[0.55,0.59]$, respectively. The line-of-sight
integration length used to project the redshift space correlation function is
$60 \mpch$. We have shifted the lowest (highest) redshift bin points to the
right (left) by a small amount for clarity. For fixed stellar mass bins, the
clustering amplitude varies very weakly with redshift. The bottom panels show
the ratio of the measured clustering of galaxies at redshifts $[0.47,0.51]$ and
$[0.55,0.59]$ to that at redshift $[0.51,0.55]$ using filled blue and open red
circles, respectively. At fixed stellar mass, the clustering does not vary
significantly with redshift.
}
\label{fig:wp_fix_mstel}
\end{figure*}

In Figure~\ref{fig:wp_fix_z}, we explore the stellar mass and redshift
dependence of the clustering signal. For this purpose, we use the galaxy
subsamples enclosed in the six color boxes in Figure~\ref{fig:Mstel_z}. Each
panel shows the dependence of the clustering signal on stellar mass selection at
fixed redshift\footnote{We use a line-of-sight integration length $\pi_{\rm
max}=60\mpch$ for projecting the redshift space correlation function for these
tests only given the limited redshift range of the data. Everywhere else in the
paper we use $\pi_{\rm max}=100\mpch$.}. This figure demonstrates that the
clustering signal varies with stellar mass at fixed redshift. Given that the
stellar mass threshold of the full sample of CMASS galaxies varies with
redshift, the stellar mass dependence of the clustering signal implies the
necessity of a proper redshift dependent modelling of the clustering if the
entire galaxy sample is used. In Figure~\ref{fig:wp_fix_mstel}, we show that
the clustering signals of the stellar mass limited subsamples do not vary
substantially with redshift. Although not a formal justification, this result
supports our assumption of a single effective redshift for modelling our
measurements. For our main analysis, we will therefore focus on subsamples A, B
and C that are defined by constant stellar mass thresholds in the redshift range
$z\in[0.47,0.59]$. 

\begin{figure*} \centering{
\includegraphics[]{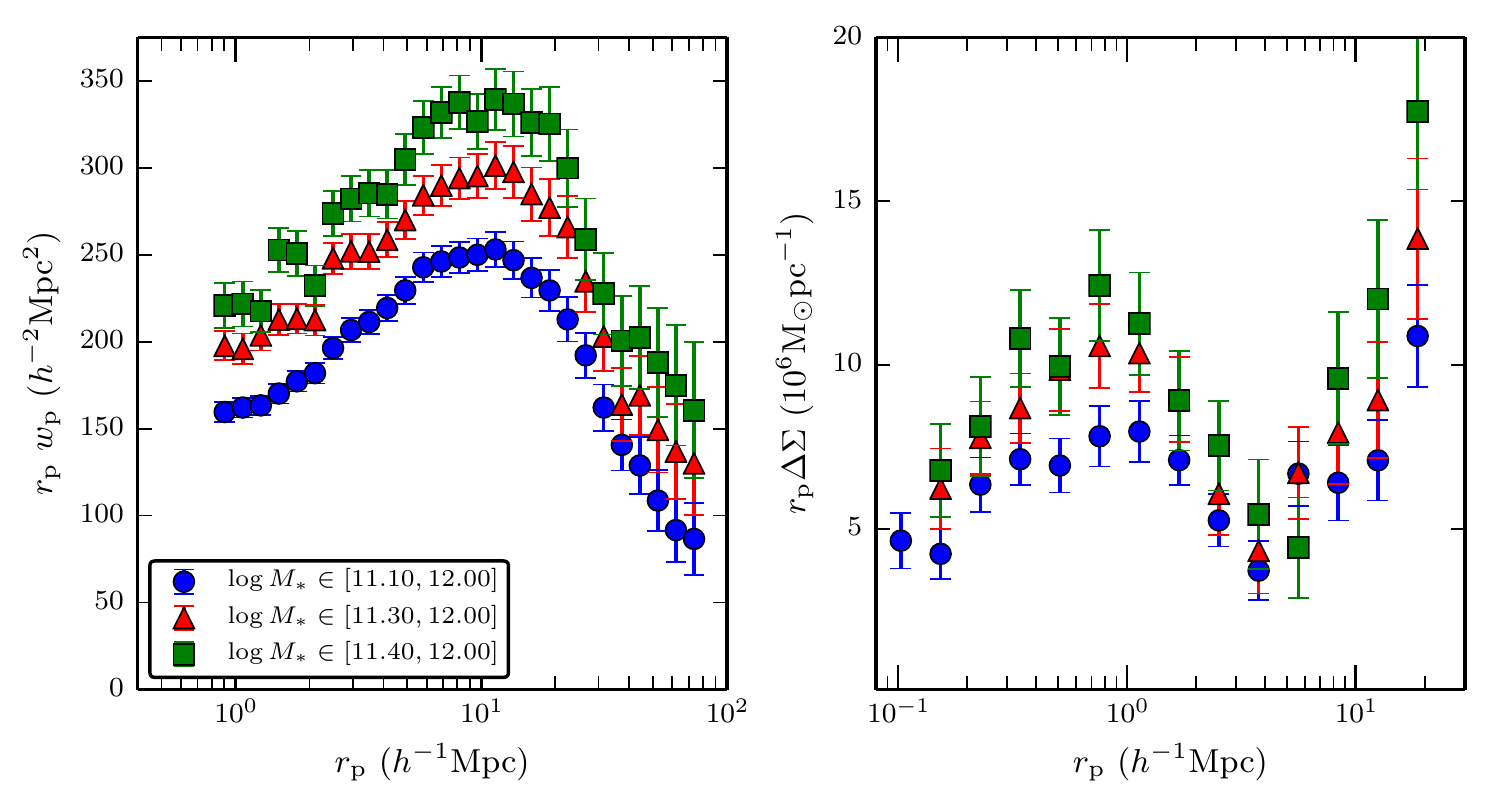}} \caption{ The clustering
(left) and the lensing (right) signal measurements for subsamples
in the redshift range $z\in[0.47,0.59]$ but with stellar mass threshold
cuts $\log M*>11.10$, $11.30$ and $11.40$ are shown using blue circles
with errors, red triangles with errors and green squares with errors,
respectively. The line of sight integration length used to project the
redshift-space correlation function is $100 \mpch$ for all the three
subsamples.  These measurements will be used to constrain the halo
occupation distribution parameters of galaxies as well as the
cosmological parameters in Paper II.}
\label{fig:cosmo_sys}
\end{figure*}

\begin{figure*} \centering{
\includegraphics[]{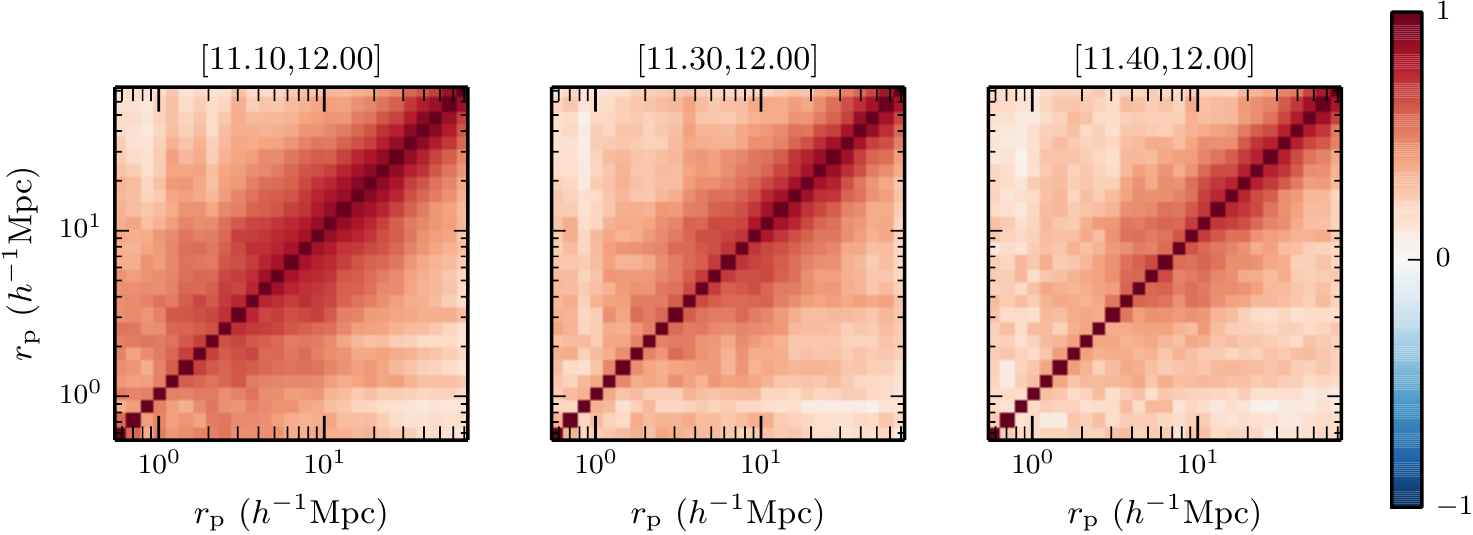}} \caption{ The
correlation matrix of the clustering measurements for the subsamples,
estimated by using 192 jack-knife samples. The covariance will be used
while calculating the likelihood of the clustering measurements given
the parameters.} \label{fig:wp_cov}
\end{figure*}

We show the
projected clustering signals for subsamples A, B and C of the CMASS galaxies in
Figure~\ref{fig:cosmo_sys}. The errorbars are obtained using the jackknife
technique, where we utilized 192 jackknife regions on the sky covering the
entire survey footprint. The cross-correlation matrix of the projected clustering
measurement in each of our subsamples (normalized to have a value of unity on
the diagonals) are shown in each of the panels of
Figure~\ref{fig:wp_cov}. The correlation coefficients are defined as
\begin{equation}
C_{ij} = \frac{{\rm Cov}(i,j)}{\sqrt{{\rm Cov}(i,i){\rm Cov}(j,j)}},
\end{equation}
where the subscripts $i,j$ denotes the radial bin index.

We observe that there is a significant covariance between the
errorbars of the measurements on large scales, and we include this
covariance while modelling the signal. The total signal-to-noise ratio of the
clustering in our fiducial subsample is $55.6$, properly accounting for the
covariance. It decreases to total signal-to-noise ratios of $48.3$ and $39.7$
for the subsamples B and C, respectively. The right panel of
Figure~\ref{fig:cosmo_sys} shows the lensing signal of our CMASS
subsample. The details of this measurement are described in the next section.


\section{CFHTLenS data and lensing measurement}
\label{sec:cfhtlens_and_lensing_measurement}

\subsection{CFHTLenS catalog}
\label{subsec:CFTHLenScatalog}
\begin{figure*}
\begin{center}
\includegraphics[width=15cm]{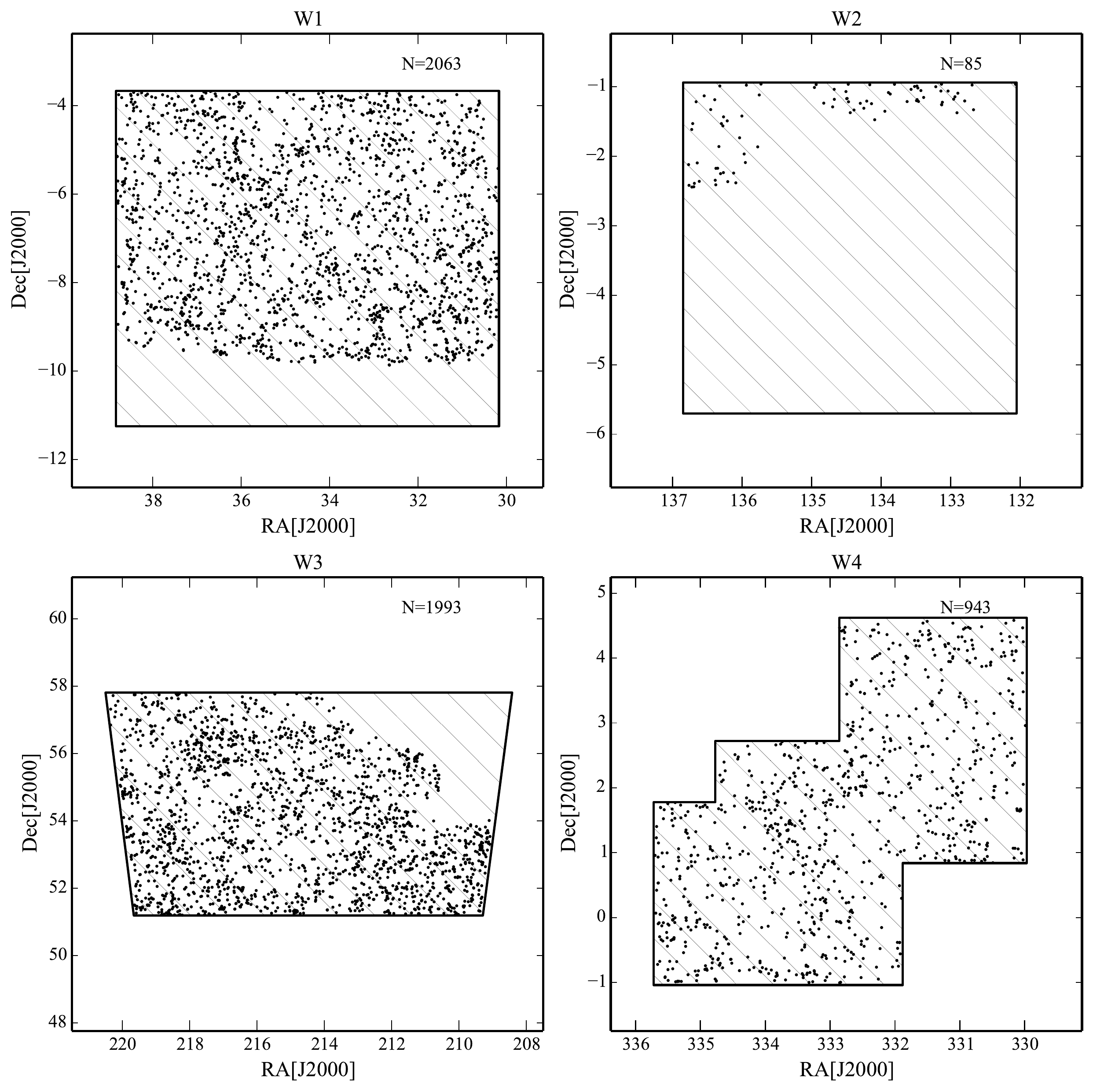} \caption{Distribution of
our fiducial CMASS galaxy subsample (subsample A) in each of the four
CFHTLenS fields as labeled at the top of each panel. The number of CMASS
galaxies in each CFHTLenS field is given in the upper right of each
panel. The hatched regions denote CFHTLenS fields.  The CMASS galaxy
subsample in this paper is selected based on their redshift and
stellar mass estimates so that the subsample constitutes approximately
volume-limited sample and physically-similar population of galaxies (see
Section~\ref{sec:BOSS_data} and Figure~\ref{fig:Mstel_z} for details).}
\label{fig:lens_galaxies}
\end{center}
\end{figure*}

For the measurements of the galaxy-galaxy lensing signal around the
subsamples of CMASS galaxies, we must measure the tangential distortion
of background galaxies. For this purpose, we rely on the deeper and
better quality imaging data from the Canada France Hawaii Telescope
Legacy Survey (CFHTLS). This information allows us to measure the
tangential distortion of background galaxies around the different subsamples of CMASS
galaxies. In particular we make use of the photometric reduction and
image shape determinations in the publicly available CFHTLenS
catalog\footnote{\url{http://www.cfhtlens.org/astronomers/data-store}}.
Unfortunately, the overlap between the CFHTLS and the DR11 BOSS fields
is limited to an area of only $\sim100$ deg$^2$. The number of CMASS
galaxies that lie within the CFHTLS footprint is $5,084$ for our fiducial
subsample A, $2,549$ from subsample B and $1,577$ for subsample C,
compared to $\sim 0.4$ million CMASS galaxies in the entire BOSS footprint.
In Figure~\ref{fig:lens_galaxies}, we show the different CFHTLS fields.  The
positions of CMASS galaxies in subsample A within these fields are indicated by
black dots.

The quantities needed for shape estimate of each galaxy image, its
ellipticity, calibration factors, and weight are provided in the
CFHTLenS catalog \citep{Heymans:2012, Erben:2013, Miller:2013}. The two
ellipticity components in the celestial coordinate system, $(e_1, e_2)$,
were estimated from the $i'$-band data of each galaxy image using the
{\it lens}fit software, which is based on a Bayesian model-fitting
method \citep{Milleretal:07} for a model with two components. The
ellipticity is defined as $e=\left(a-b\right)/\left(a+b\right)$, where
$a$ and $b$ are the major and minor axes of the ellipse,
respectively. Using a shear recovery test based on galaxy image
simulations, the CFHTLenS team also provided calibration factors that
are a function of galaxy size and detection signal-to-noise ratio, so
that the input shear for simulated galaxy images is recovered to the
desired accuracy after application of these factors. The calibration
factors consist of the shear multiplicative bias factor $m$, which is
commonly applied to both $e_1$ and $e_2$, and the additive term $c_2$,
which is applied to $e_2$ alone. The shear correction is greater for a
galaxy with small signal-to-noise ratio and small scale-radius
(size). We will describe the details of the shear calibration scheme in
Section~\ref{subsec:lensing_signal_calculation}. The inverse-variance
weight for each galaxy is defined by the variance that is estimated from
the intrinsic galaxy ellipticity and the measurement error due to photon
noise \citep[for details, see][]{Miller:2013}.

Photo-$z$ for each source galaxy were estimated with the {\it BPZ} code
\citep{Benitez:2000, Coe:2006} by using PSF-matched photometry that aims
at measuring aperture photometry for the same physical part of each
galaxy in different passbands (for details, see
\citealt{Hildebrandt:2012}). The BPZ code provides a probability
distribution function (PDF) of the photo-$z$ estimate for each galaxy
(hereafter $P(z)$). We make use of the full information of $P(z)$ when
computing lensing signal.

To make a reliable lensing measurement, we use the following catalog of
source galaxies. First, we discard galaxies that have the flag {\tt
MASK} $> 1$ indicating masked objects. We use the galaxies that have the
ellipticity weight {\tt weight} $>0$, and the ellipticity fitting flag
{\tt fitclass} $=0$, which indicates that the shape is reliably
estimated. We do not apply any cut to magnitude or signal-to-noise
ratio, i.e., we use all the faint galaxies as long as the above
conditions are satisfied. Although the faint galaxies are highly
downweighted and largely corrected for the calibration factors, they
contribute to the lensing signal and slightly increase the
signal-to-noise ratio.

\subsection{Galaxy-galaxy lensing measurements}
\label{subsec:lensing_signal_calculation}

Galaxy-galaxy weak lensing measures a coherent distortion of source
galaxy shapes due to all matter around lens galaxies, including dark
matter \citep[see][and references therein]{Mandelbaum:2013}.  The
lensing signal is only statistically measurable and can be estimated by
stacking tangential component of source galaxy ellipticities with
respect to the position of lens galaxy, for all the pairs of lens and
source galaxies in each circular annulus.  The lensing distortion
profile probed in this way is expressed in terms of the projected
surface mass density profile of the average mass distribution around the
lens galaxies:
\begin{equation}
\label{eq:shear_signal}
\gamma_t(R) = \frac{\Delta\Sigma(R)}{\Sigma_{\rm{cr}}} =
 \frac{\bar{\Sigma}(<R) - \Sigma(R)}{\Sigma_{\rm{cr}}},
\end{equation}
where $R$ is the projected separation between the source and lens
galaxies at the redshift of each lens galaxy, $\Sigma(R)$ is the
projected mass density profile at radius $R$, $\bar{\Sigma}(<R)$ is the
average mass density within a circle of radius $R$, and $\Sigma_{\rm
cr}$ is the critical surface mass density.  A spectroscopic redshift for
each CMASS galaxy, $z_\rml$, enables an estimation of the projected
radius from the observed angle separation $\Delta\theta$ via
$R=d_A(z_\rml)\Delta \theta$, where $d_A(z_\rml)$ is the comoving
angular diameter distance to the lens galaxy.  The critical density
$\Sigma_{\rm cr}$ for lens and source galaxies at redshifts $z_\rml$ and
$z_\rms$, respectively, is defined as
\begin{equation}
\Sigma_{\rm{cr}}^{-1}\left(z_\rml, z_\rms\right) = 
\frac{4\pi G}{c^2}\frac{d_A(z_\rml)d_A(z_\rml,z_\rms)(1+z_\rml)^2}{d_A(z_\rms)}.
\end{equation}
Here $d_A(z_\rms)$, $d_A(z_\rml)$ and $d_A(z_\rml, z_\rms)$ are the
 angular diameter distances for the source-lens system. The factor of
 $(1+z_\rml)^{2}$ arises from our choice of comoving
 coordinates. Another component of shear, $\gamma_\times$, which is a
 $45^\circ$ rotated component from the tangential shear, should be
 statistically consistent with zero for weak gravitational lensing (but
 potentially nonzero for shape distortions due to systematic
 errors). Hence we can use the measured $\gamma_\times$ as a monitor of
 a possible residual systematics in the lensing measurement.

For each pair of lens and source galaxies, we compute the tangential
ellipticity component using
\begin{equation}
\label{eq:tangential_component}
e_{t} = -e_1 \cos2\phi - e_2 \sin2\phi,
\end{equation}
where $\phi$ is defined as the angle measured from right ascension
direction to a line connecting the lens and source galaxies at source
galaxy position. Using spherical trigonometry, the angle $\phi$ is given
in terms of the galaxy positions ($\alpha,\delta$) as
\begin{eqnarray}
\cos\phi&=&
 \frac{\cos\delta_1\sin(\alpha_2-\alpha_1)}{|\sin\theta\,|}\nonumber\\
\sin\phi&=&\frac{-\sin\delta_1\cos\delta_2+\cos\delta_1\cos(\alpha_2-\alpha_1)
\sin\delta_2
}{|\sin\theta\,|},
\end{eqnarray}
where the angles with subscripts ``1'' or ``2'' denote the coordinate
components for
the lens or source galaxies, respectively, and $\theta$ is the
separation angle between the galaxies on the sphere:
\begin{equation}
\cos\theta = \sin\delta_1 \sin\delta_2 + \cos\delta_1 \cos\delta_2
 \cos\left(\alpha_1 - \alpha_2\right).
\end{equation}
The $45^\circ$ rotated component, $e_\times$, is also similarly
computed.

While the lens equation (Eq.~\ref{eq:shear_signal}) is for a single
source redshift $z_\rms$, we take into account the uncertainty in the
posterior distribution of the source galaxy photometric redshift. This
uncertainty dominates the small uncertainty in the spectroscopic
redshift estimate of lens CMASS galaxies. 

We follow the method of \cite{Mandelbaum:2013}, and estimate the average 
projected mass density profile, $\Delta\Sigma(R)$ in
Eq.~\eqref{eq:shear_signal}, as
\begin{equation}
\label{eq:delta_sigma_from_data}
\Delta\Sigma\left(R\right) = \frac{\sum_{\rm ls} w_{\rm ls}e_{t}^{(\rm ls)}\left[\left\langle\Sigma^{-1}_{\rm cr}\right\rangle^{(\rm ls)}\right]^{-1}}{\left(1+K\left(R\right)\right)\sum_{\rm ls} w_{\rm ls}},
\end{equation}
where the summation runs over all the pairs of source-lens galaxies separated
by the projected radius $R$ to within a given bin width, the superscripts
``$l$'' or ``$s$'' stand for lens or source galaxies, respectively, and
$\left\langle\Sigma^{-1}_{\rm cr}\right\rangle^{(\rm ls)}$ is the critical density
averaged with the photo-$z$ PDF for each source-lens pair, defined as
\begin{eqnarray}
\label{eq:sigma_inv_ave}
 \left\langle\Sigma^{-1}_{\rm cr}\right\rangle^{(\rm ls)}
 \equiv
 \frac{
 \int_0^{\infty}\!dz_\rms \Sigma_{\rm cr}^{-1}\left(z_\rml, z_\rms\right) P(z_\rms) 
dz_\rms}
{
\int_0^{\infty}\!dz_\rms P_\rms(z_\rms)}.
\end{eqnarray}
Here the integration is performed from $z=0.025$ to $z=3.475$ with
the interval of $\Delta z=0.05$, following the full $P(z)$ information from the CFHTLenS
catalog. We set $\Sigma_{\rm cr}^{-1}(z_\rml,z_\rms)=0$ at
$z_\rms<z_\rml$ in the above calculation. 
The above equation
automatically corrects for a possible dilution of the lensing signal
caused by non-vanishing probability that a source galaxy is at
$z_\rms<z_\rml$. The weight $w_{\rm ls}$ is defined as
\begin{equation}
\label{eq:weight}
w_{\rm ls} = w_\rml w_\rms 
\left(
\left\langle\Sigma^{-1}_{\rm cr}\right\rangle^{(\rm ls)}\right)^2,
\end{equation}
where $w_\rml$ is defined by Eq.~\eqref{eq:lens_weight_cmass}, $w_\rms$ is the
weight given by the CFHTLenS catalog described in Section
\ref{subsec:CFTHLenScatalog}, and the factor
$\left(\left\langle\Sigma^{-1}_{\rm cr}\right\rangle^{(\rm ls)}\right)^2$
downweights pairs that are close in redshift and therefore are
inefficient in weak lensing and vice versa. The overall factor
$(1+K(R))$ is introduced in Eq.~\eqref{eq:delta_sigma_from_data} as recommended in
\cite{Miller:2013}. This 
factor corrects for a multiplicative shear bias, and is implemented
after the stacking average, rather than on a per-galaxy basis. The
calibration factor is calculated as
\begin{equation}
1+K(R) = \frac{\sum_{\rm ls}w_{\rm ls}(1+m^{(\rm ls)})}{\sum_{\rm ls} w_{\rm ls}},
\end{equation}
where $m^{(ls)}$ is the multiplicative bias factor defined in \cite{Miller:2013}.

For a radial binning for the lensing profile, we set the innermost bin
to $R_{\rm min}=0.0245\mpch$ and employ logarithmically-spacing binning
given by $\Delta R/R=0.4$ (about 6 bins in one decade of radial bin
spacing).

\subsection{Correcting Lensing Systematic Errors with Random Catalogs}
\label{subsec:systematic_tests} The shear profile measured according to
the method given in the preceding section might still be contaminated by
residual systematic effects inherent in the data. One possible
systematic error is a dilution of the lensing signal caused when
including ``source'' galaxies, which are actually physically associated
with the lens galaxy (or its halo), into the stacking analysis.  Another
one is a possible residual systematic in the shape measurement, e.g.,
caused by an imperfect correction of optical distortion across the field
of a camera. In this section, following the method in
\cite{Mandelbaum:2005}, we use random catalogs provided by the
SDSS-III/BOSS collaboration \citep{Anderson:2014}, which is
appropriately downsampled to match our subsamples as described in
Section~\ref{sec:clustering}, to test and correct for these systematic
effects. We use random catalogs which consist of 100 times more
points than the number of CMASS galaxies, and divide them into 100
realizations to estimate the systematic uncertainties in the following
measurements.

\subsubsection{Boost Factor}
\label{subsec:boost_factor}

If some of the galaxies in the source sample are physically associated with a
lens galaxy, they will dilute the lensing signal (because
they are not lensed). For instance, this is the case if source galaxies
are in the same halo of a lens galaxy. This contamination can be
estimated by searching for an excess in the number counts of source
galaxies in the region of lens galaxies compared to the random
distribution. To study the possible excess, we can use the random
catalogs that are randomly distributed on the sky, but are generated
mimicking the redshift distribution of CMASS lens galaxies. Thus, taking
into account the weights, we estimate the {\it boost} factor defined as
\begin{equation}
\label{eq:boost_factor} B(R) = \frac{\sum_{\rm ls} w_\rml w_\rms
\left(\left\langle\Sigma^{-1}_{\rm cr}\right\rangle^{(\rm ls)}\right)^2
\left/ \sum_\rml w_\rml \right.}{\sum_{rs} w_\rmr w_\rms
\left(\left\langle\Sigma^{-1}_{\rm cr}\right\rangle^{(rs)}\right)^2
\left/ \sum_\rmr w_\rmr \right.},
\end{equation}
where the superscript ``$r$'' stands for random catalogs, and $w_{r}$ is
the weight for random-lens, at each projected radius $R$. For an ideal
source catalog, $B(R)=1$, while $B(R)>1$ if there is a contamination by
physically-associated source galaxies. 

\begin{figure*}
\begin{center}
\includegraphics[width=8cm]{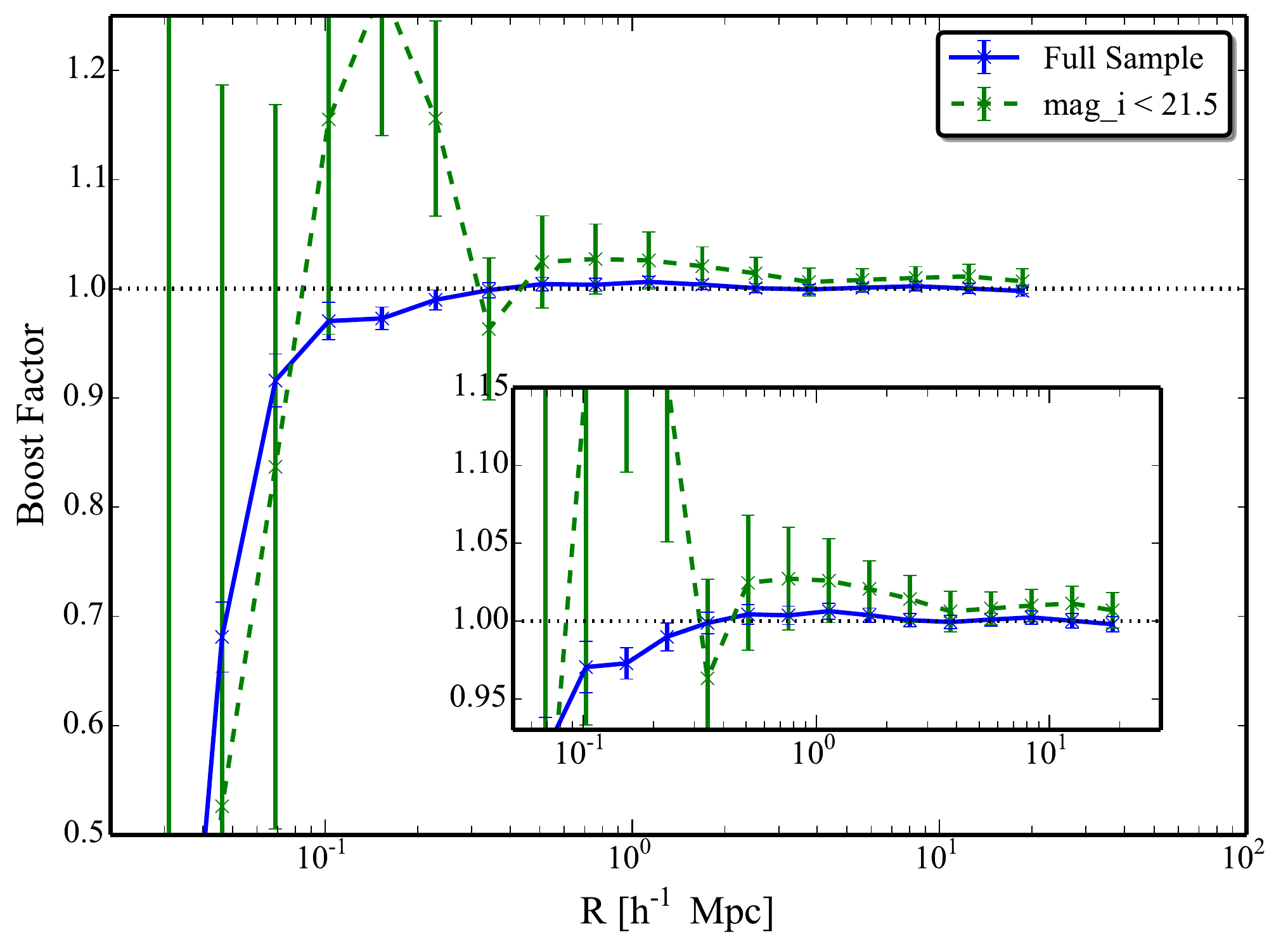}
\includegraphics[width=8cm]{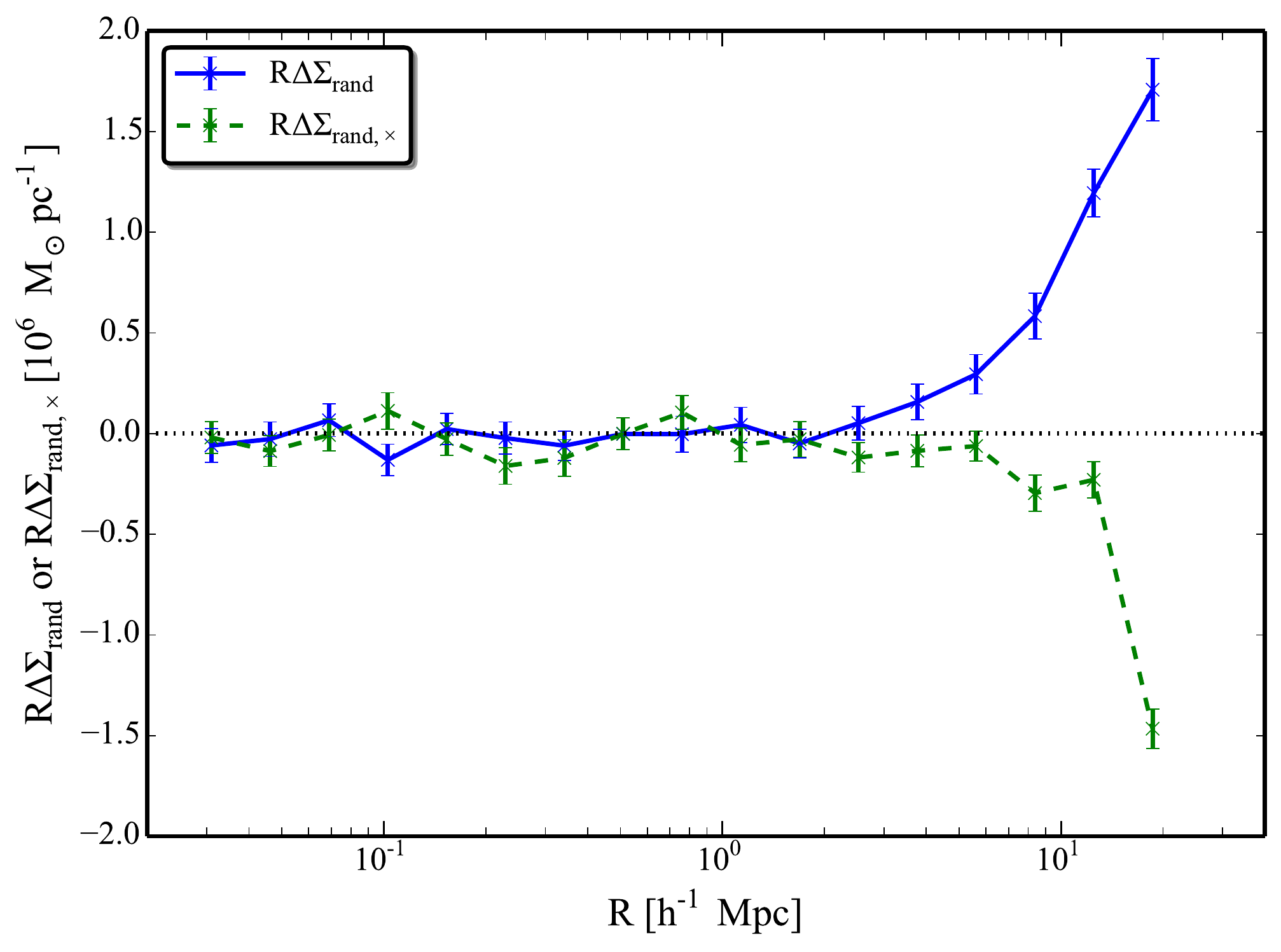} 
\caption{{\it Left panel}: A ``boost factor'' $B(R)$, which measures an
excess or deficiency in the number of the CMASS-lens and CFHT-source
galaxy pairs compared to random point and CFHT-source pairs, where the
random catalogs are generated mimicking the redshift distribution of
 our subsample A.
 The errorbars are estimated from the standard deviations of the
results for 100 random catalogs. The $x$-axis is the projected comoving
radius between the source and lens (or random point) pairs. The expanded
plot is shown in the inset at lower right.  The solid curve denotes the
boost factor for our fiducial subsample of CFHTLenS galaxies, showing that
the boost factor is consistent with unity except at small radii $R\simlt
100~\kpch$ (see text for discussion). For comparison, the dashed curve
is the boost factor for the brighter sample with $\mbox{\tt
mag\_i}<21.5$; $B(R)>1$ at radii $R\simlt 1~\mpch$, indicating that some
of the CFHTLenS galaxies are physically associated with CMASS galaxies,
such as galaxies in the same host halos. {\it Right panel:} The
lensing signals for the CMASS random catalogs for our subsample A,
which are measured by stacking the tangential (the blue curve) or
$45^\circ$ rotated (the green curve) components of the CFHTLenS galaxy
ellipticities around the random points. Since the random points are
randomly distributed on the sky, the signals should vanish within the
errorbars if the shape measurement is not contaminated by residual
systematic errors. The results show some deviation from zero at the
large radii, $R\simgt 10~\mpch$, indicating residual systematic errors
(see text for discussion).} \label{fig:systematics}
\end{center}
\end{figure*}

The left panel of Figure~\ref{fig:systematics} shows the boost factor we
measured for our subsample A, where the errorbars are estimated
from the scatters of random realizations.  The small scales at $R\simlt
100\kpch$ display $B(R)<1$, implying that the number of source galaxies
behind the CMASS lens galaxies is smaller than the random distribution,
which is not expected from physically-associated source galaxies. A
possible origin of $B(R)<1$ is that a contamination from light of
foreground CMASS galaxies reduces the efficiency of detecting source
galaxies. The CMASS galaxies are indeed the brightest galaxies in the
redshift range in the SDSS catalog, and can be too bright (perhaps
causing saturated images) for a much deeper survey such as
CFHTLenS. Given that these details rely on the detection of objects in
the CFHTLenS {\it lens}fit processing, we are not in a position to
investigate this issue further. Another possibility is
sky-subtraction. The size of sky mesh is
128~pixels\footnote{priv. comm. with T. Erben.}, which corresponds to
$\sim100 \kpch$ at the redshift of CMASS galaxies. Therefore an
over-subtraction of the sky might be a possible origin of $B(R)<1$ at
the small radii. Magnification can also affect the observed counts of
galaxies near bright objects. \cite{Duncan:2014} showed a deficit
is expected for the similar magnitude depth\footnote{On the other hand,
they showed an excess as expected for a bright sample {\tt mag\_i}
$<21.5$.}. Because of the difficulty in identifying the origin of
possible systematic errors on small scales, in the following analysis we
do not use data at scales smaller than $100 \kpch$ (we use
different cuts for other subsamples, which are described later). On the
other hand, as shown on the plot, the boost factor is consistent with
unity at scales larger than $100 \kpch$, meaning that a contamination of
physically-associated galaxies with CMASS galaxies is negligible. Thus
we need not adopt the boost factor correction for the CMASS lensing at
$R>100\kpch$. For comparison, we also plot in
Figure~\ref{fig:systematics} the boost factor measured for brighter CFHT
galaxies with {\tt mag\_i} $<21.5$. In this case, $B(R)>1$ at $R\simlt
1\mpch$, comparable with a virial radius of massive halos. This result
implies that some of the bright galaxies are physically associated with
CMASS galaxies, even if their photo-$z$'s are higher than the CMASS
galaxy redshift.

\subsubsection{Testing Lens Signal of Random Catalogs}
\label{sec:random}

We can also use the random catalogs of CMASS galaxies for testing
possible residual systematic errors in the shape measurement of CFHTLenS
source galaxies. Since the random points are randomly distributed on the
sky, we should not detect any coherent tangential distortion of CFHTLenS
galaxy shapes around the random points, if the shape measurement is not
contaminated by systematic errors. This test can be done by using the
random points instead of CMASS galaxy positions in
Eq.~\eqref{eq:delta_sigma_from_data}.

The right panel of Figure~\ref{fig:systematics} shows the tangential and
$45^\circ$ rotated distortion profile measured from the random catalogs
for our subsample A. While the distortion is consistent with zero
within the errorbars at $R\simlt 3\mpch$, the larger radii show an
increasing deviation from zero. Thus the larger radius scales indicate
residual systematic errors in the shape measurement. The projected
radius $R=10~\mpch$ for the CMASS mean redshift $z\simeq 0.53$
corresponds to about $1^\circ$, comparable with the field-of-view of
CFHT/MegaCam. Thus the systematic errors are likely due to an imperfect
PSF correction in the galaxy shape measurement, or more precisely an
imperfect correction of the optical distortion of the camera, which
tends to cause a tangential or radial pattern of the PSF ellipticities
in the edge of the field of view \citep{Hamanaetal:13}.  Since the
source and random-point pairs of larger radii are preferentially
sensitive to a coherent PSF anisotropy in the edge of field-of-view, the
result indicates such a residual systematic error. In the following
analysis, we correct for the lensing signals of CMASS galaxies by
subtracting the random signal in Figure~\ref{fig:systematics} from the
measured signal. As described in \cite{Mandelbaum:2005} \citep[see
also][]{Mandelbaum:2013}, this correction rests on the assumption that
the distribution of lens galaxies is uncorrelated with residual
systematics in the shape measurements. This assumption holds in our
analysis because the lens catalog and the shape measurements are taken
from completely different datasets, the SDSS and CFHTLenS data.  The
size of correction is consistent with zero at small radii, and increases
to $10$-$15\%$ of the measured signals at large radii at $R\sim10\mpch$.

\subsection{CMASS Galaxy-galaxy Lensing Signal}
\label{subsec:CMASS_galaxy_signal}

\begin{figure}
\includegraphics[width=8cm]{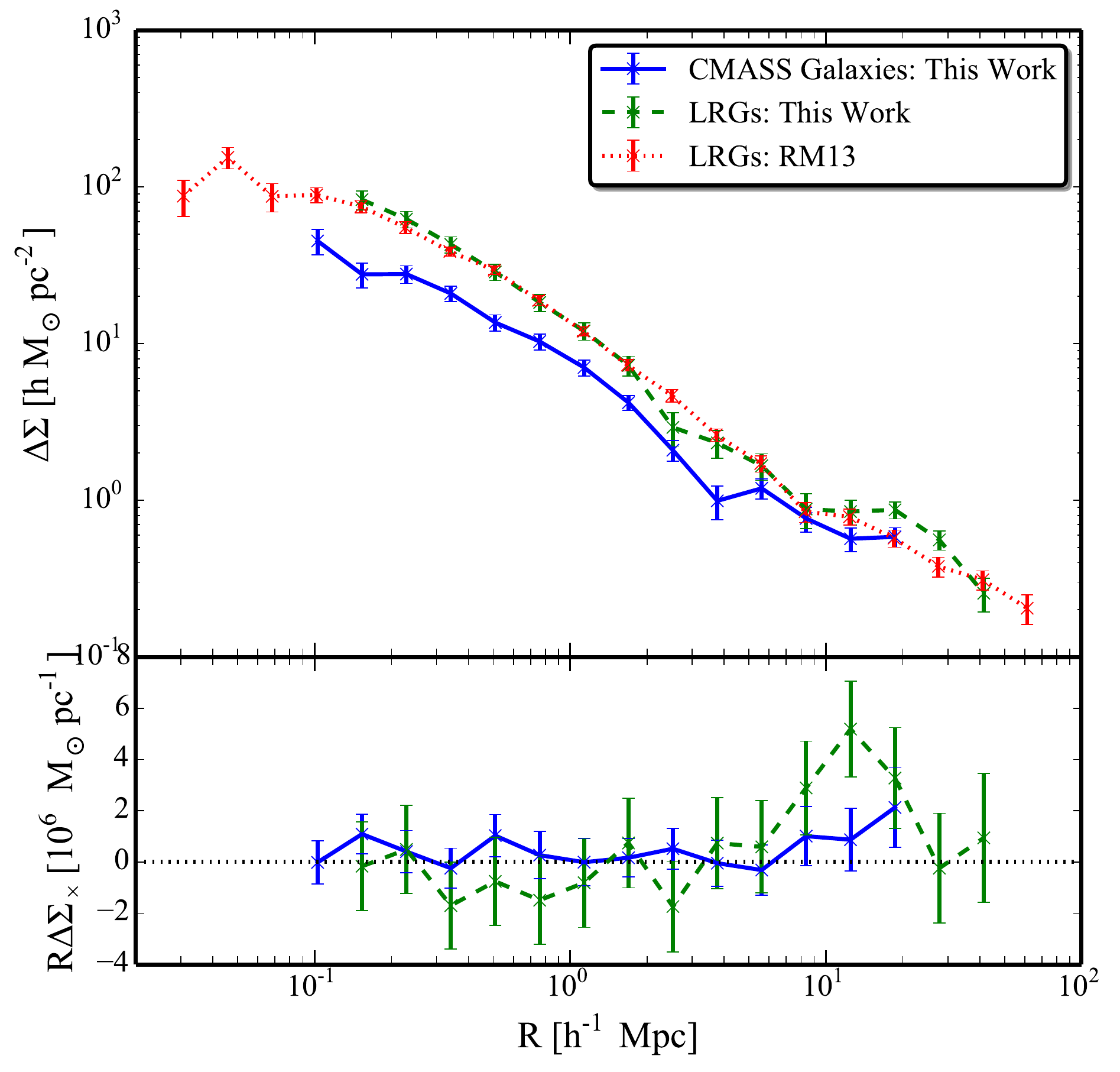} \caption{The solid curves
show the average surface mass density profile measured from the CMASS
and CFHTLenS galaxy-galaxy weak lensing, where the errorbars are
estimated from 100 random catalogs of CMASS galaxies.  In this
measurement we corrected for the possible contamination of residual
systematic errors in Figure~\ref{fig:systematics}; we subtracted the
lensing signals of the random catalogs from the measured CMASS lensing
signals at each radial bin. For comparison, the dashed curve shows the
lensing profile for the SDSS luminous red galaxies (LRGs), measured from
the CFHTLenS catalog using the same measurement procedures. The dotted
curve is the LRG lensing measured from the SDSS catalog, which shows an
excellent agreement with the CFHTLenS measurement within the
errorbars. The lower panel presents the measurements of non-lensing mode
for the CMASS and LRG catalogs. } \label{fig:lens_signal}
\end{figure}

In Figure~\ref{fig:lens_signal}, we present the CMASS galaxy-galaxy
lensing signals for our subsample A, after correcting for the
systematic errors as we described above. As can be found from the lower
panel, the signal in the $45^\circ$ rotated component is consistent with
zero over all the radii we consider. The upper panel shows the
tangential distortion, which shows the expected trends of decreasing
signal as the projected radius increases. The cumulative signal-to-noise
ratio is about $26$. For comparison, we show the lensing signal of the
SDSS LRG catalog using the CFHTLenS source catalog and the same lensing
measurement procedures (for details of the LRG lensing measurement, see
Appendix~\ref{app:LRG_lensing_signal}). The CMASS galaxy lensing signal
is about $40\%$ smaller than the LRG lensing signal, suggesting that the
CMASS galaxies preferentially reside in less massive halos. As a
verification of our lensing measurement, we also show the LRG signal
measured from the independent data, the SDSS source catalog, in
RM13. The CFHTLenS and SDSS LRG lensing signals are in excellent
agreement with each other. To be more quantitative, the inverse-variance
weighted ratio of the two LRG signals from these different surveys with
completely independent shape measurements and photo-$z$, averaged over
all the radial bins, is $1.006\pm0.046$. We note that the LRG lensing
measurement is performed under the cosmological model of $\Omega_m=0.25$
and $\Omega_\Lambda=0.75$ to make it consistent with the measurement in
RM13.

We comment on a possible systematic bias in the CMASS lensing signal
that may be caused by a residual bias in the photo-$z$ estimates of
source galaxies, to the level of $\delta z\sim 0.02$, due to photo-$z$
outliers as claimed in \cite{Erben:2013}.  We checked that even shifting
the photo-$z$ PDF, $P(z)$, for all the source galaxies by $\delta z=\pm
0.02$ in the analysis causes only a few percent shift in the lensing
signal. Hence we ignore the possible photo-$z$ bias in the following
results.

\begin{figure*} \centering{
\includegraphics[]{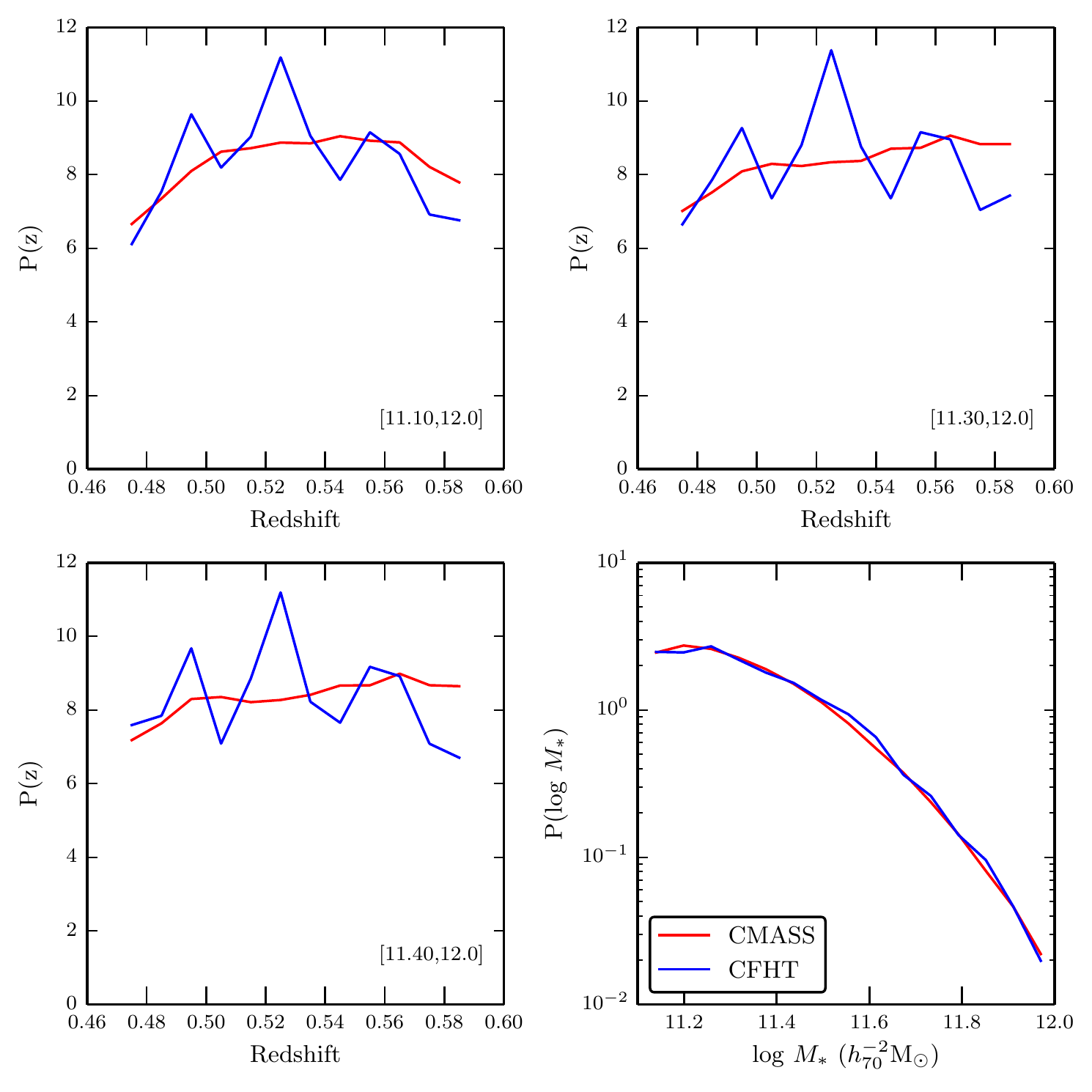}}
\caption{
The comparison of the redshift distributions of galaxies in the CFHTLS fields
(blue solid line) and those in the CMASS fields (red solid line) for the three
stellar mass subsamples we use for our analysis are shown in the
top left, top right and bottom left panels. The bottom right panel shows a
similar comparison but for the stellar mass distribution of galaxies. The
stellar mass distribution of the CFHT subsample appears fairly representative of
that of the entire CMASS population. The redshift distribution shows noisy
deviations due to structures in the CFHT regions.
}
\label{fig:compare_cfht_cmass}
\end{figure*}

Another systematic uncertainty might come from the difference
between the CMASS galaxy subsample in the entire BOSS region and
that in the CFHTLS region. We compare the probability distribution of
the stellar mass, $P(\log M_*)$, and that of the redshifts, $P(z)$, for
our subsamples in these regions in the panels of
Figure~\ref{fig:compare_cfht_cmass}.  Although the stellar mass
distributions in the CFHT region is not particularly special, the
redshift distributions shows some noticeable differences, presumably due
to the structures in the CFHT field. We assessed the systematics of the
differences in the redshift distribution by including a weight $w_{\rm
sys}(z)=P^{\rm CMASS}(z)/P^{\rm CFHT}(z)$, where $P^{\rm CMASS}(z)$
and $P^{\rm CFHT}(z)$ is the spectroscopic distribution in the entire
CMASS region and that in the CFHTLS region, respectively, when calculating the weak
lensing signal. The differences in the lensing signal when including
this weight are of the order $2$ percent, much smaller than our
errorbars on the lensing signal.  We therefore conclude that this weight
is not needed for the analysis, and that the subsample of CMASS
galaxies in the CFHT fields is sufficiently close to a fair sample for
our purposes.

\begin{figure*}
\centering
\includegraphics[width=16cm]{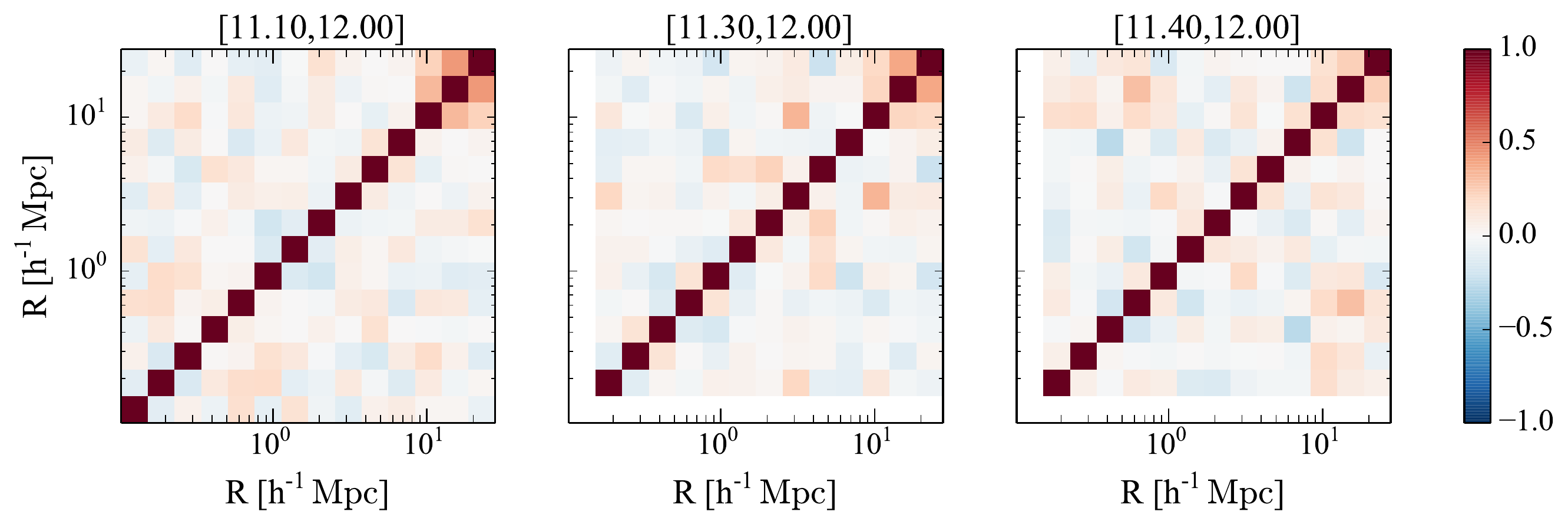}
\caption{Correlation coefficients of the error covariance matrix for the
 CMASS galaxy weak lensing signal for our subsamples,
 where the covariance matrix is computed from the 100 random
catalogs.} \label{fig:cmass_covariance}
\end{figure*}

\begin{figure}
\centering
\includegraphics[width=8cm]{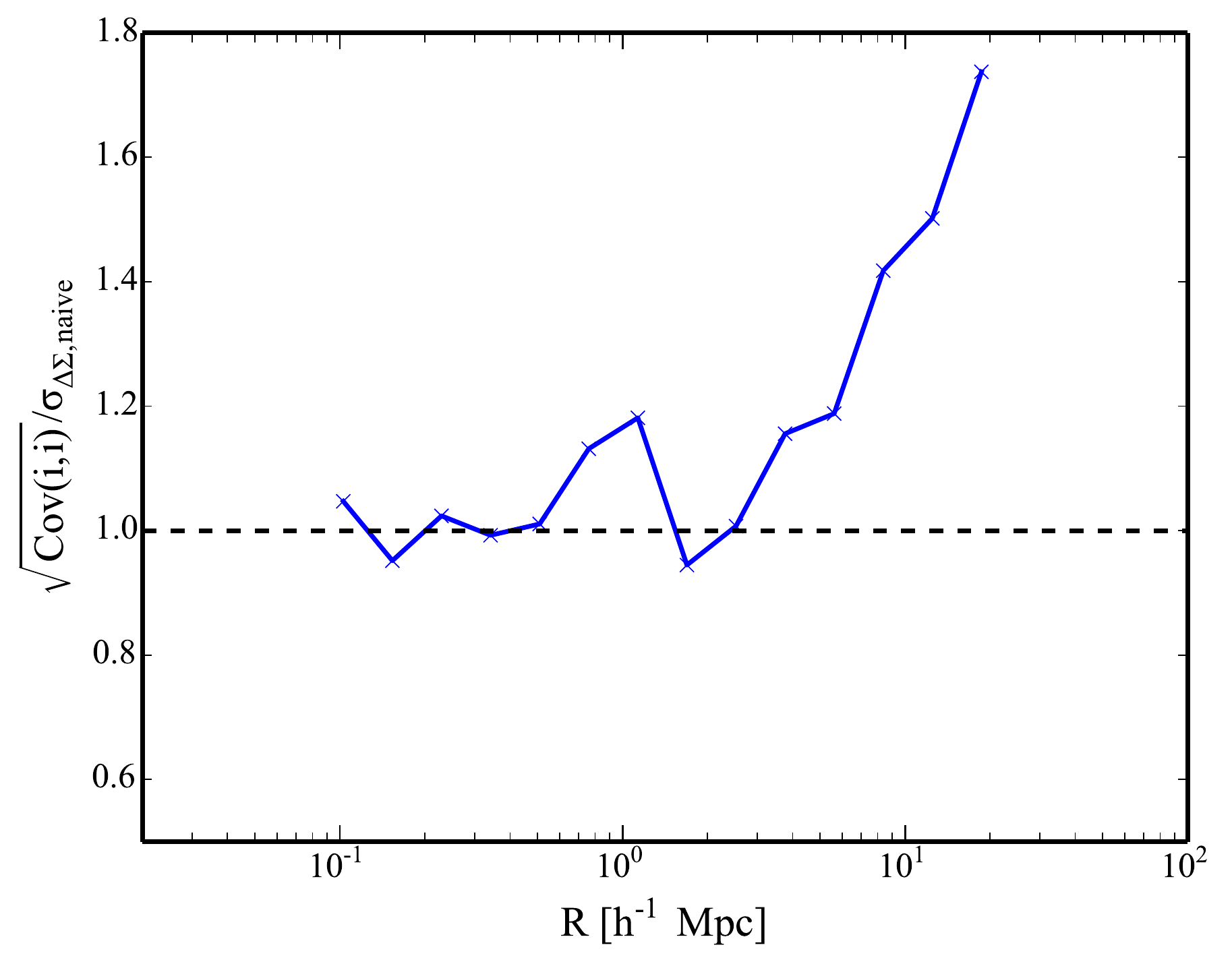} 
\caption{The square
root of the diagonal covariance elements relative to the
naively-expected shape noise, where the latter is estimated by scaling
the shape noise by $\sqrt{N_{\rm pair}}$ (the total number of
source-lens galaxy pairs at each radial bin). To estimate the shape
noise expectation, we properly take into account the weights.  At
$R\simlt $ a few $\mpch$, the shape noise is a dominant source of the
errors, while the projection effect and the correlated shape noise
become significant at larger radii.} \label{fig:cmass_error_comparison}
\end{figure}

To construct a model interpretation of the CMASS lensing signal, we need
to compute the error covariance matrix of the lensing profile.  There
are two sources of the statistical errors. First, since the number of
source galaxies used is finite, the intrinsic shape noise
contributes to the errors. In fact this is a dominant source of the
errors over the range of radial scales we consider. The shape noise is
naively expected to scale as $\sigma_e/\sqrt{N_{\rm pair}}$, where
$\sigma_e$ is the rms intrinsic ellipticity combined with the rms
ellipticity measurement error, per component, and $N_{\rm
pair}$ is the number of source-lens pairs used in the lensing
measurement of a given radius. However, the scaling does not hold for
large radii, because the same source galaxies are used multiple times as
the stacking annuli of such large radii overlap for different lens CMASS
galaxies. The other noise source arises from a projection effect:
large-scale structure along the same line-of-sight to the CMASS lens
galaxy, at different redshifts, causes statistical scatters in the
distortion of CFHTLenS galaxies. We estimate these contributions by
computing the covariance matrix of lensing signals from the 100
random catalog realizations. The random catalog enables us to estimate
both the covariance contributions of the shape noise and the projection
effect. We note that we cannot use the jackknife method for the
covariance estimation of the lensing measurements. Unlike for the BOSS
clustering measurements, the area of the CFHTLS region which overlaps
with BOSS is too small to have a sufficiently large enough resampling of
the CFHTLenS regions for the jackknife method (especially for the large
separation radii). The left panel of Figure~\ref{fig:cmass_covariance}
shows the correlation coefficients of the covariance matrix for
subsample A.  The large separation radii display a strong correlation
between neighboring bins. Figure~\ref{fig:cmass_error_comparison}
shows the ratio of the diagonal covariance elements to the
naively-expected shape noise error, where the shape noise expectation is
computed taking into account the weights.  At larger scales, the ratio
is significantly greater than unity, meaning that the projection effect
and the correlated shape noise become significant at these radii.

The lensing signals for the
three subsamples A, B and C of CMASS galaxies are shown in the right panel of
Figure~\ref{fig:cosmo_sys}.  We perform the same analysis
for subsamples B and C, such as the systematic tests by boost factor and
correction for imperfect PSF modeling by random signals.  For subsamples
B and C we find that the boost factor is significantly smaller than one
at scales below $150\kpch$, and thus discard the lensing signal at
these scales. As the stellar mass threshold increases, the amplitude of
lensing signal becomes larger. Correlation coefficients of subsamples B
and C are shown in the middle and right panel of
Figure~\ref{fig:cmass_covariance}, respectively.

\begin{figure}
\centering
\includegraphics[width=8cm]{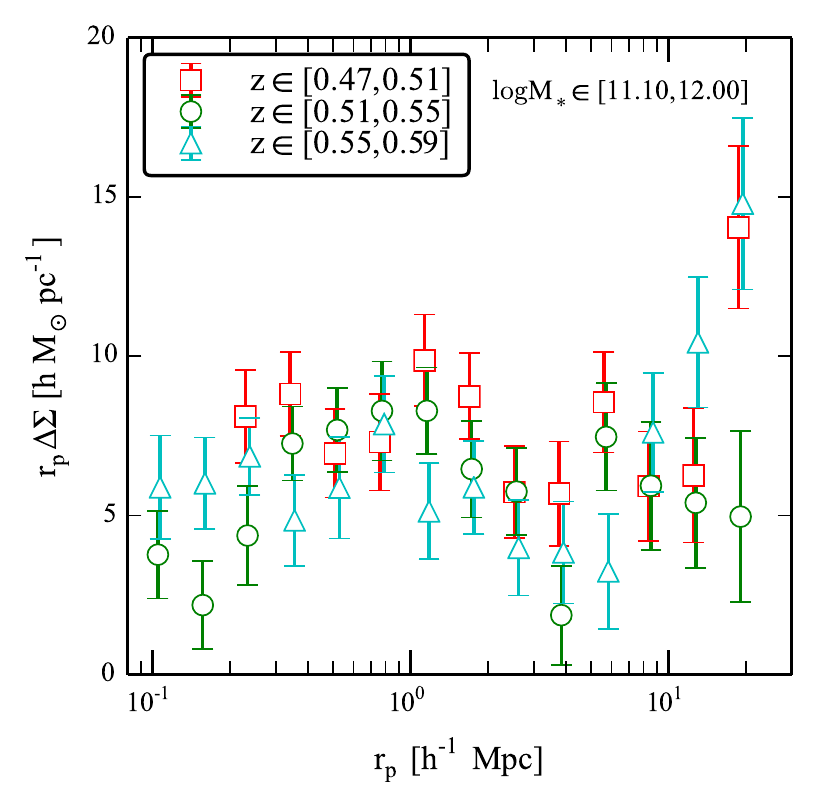} 
\caption{The lensing signals of redshift subsamples at fixed stellar
 mass threshold $\log M_\ast \in [11.10, 12.00]$. There is no
 substantial variation with redshift.}
 \label{fig:cmass_lensing_redshift_subsamples}
\end{figure}

We also explore the redshift dependence of the lensing signal. In
Figure~\ref{fig:cmass_lensing_redshift_subsamples}, we show lensing
signals for three redshift subsamples with the lowest stellar mass
threshold. The lensing signals do not vary substantially with redshift,
similar to the behaviour of the clustering signal.

\section{Discussion and conclusions}
\label{sec:conclusion} 

In this paper, we have shown the clustering and lensing
measurements of the SDSS-III/CMASS galaxies at $z\simeq 0.53$. In our
analysis, we constructed a subsample of galaxies so that it constitutes
an approximately stellar mass limited sample ($\log M_\ast/h_{70}^{-2}$
in $[11.10,12.0]$) over the redshift range $z\in [0.47,0.59]$, with
approximately constant number density (subsample A).  This subsample of
galaxies was used to measure the projected clustering signal with a
total signal-to-noise ratio of 56 for scales $0.85~h^{-1}{\rm Mpc}
\simlt r_p \simlt 80~h^{-1}{\rm Mpc}$ (see
Figure~\ref{fig:cosmo_sys}). We then made use of the publicly available
galaxy shape and photometric redshift catalogs compiled by the CFHTLenS
collaboration based on deeper, higher quality imaging data from the
CFHTLS.  This imaging catalog has an overlap of a mere 105 deg$^2$ with
the BOSS footprint, but it allows measurement of the weak gravitational
lensing signal of the BOSS galaxies, selected in the same way to our
fiducial subsample, with a signal-to-noise ratio of 26 for
$0.1~h^{-1}{\rm Mpc} \simlt r_p \simlt 20~h^{-1}{\rm Mpc}$.  Thus we
showed significant detections for both the clustering and weak lensing
signals for the BOSS CMASS galaxies at the highest redshift for such a
joint analysis, $z\sim 0.5$.

To test for possible systematics arising from our sample selection
we also studied the clustering and lensing signals for two other
subsamples within the same redshift range, but with higher thresholds in
stellar mass, $\log M_\ast/h^{-2}_{70}M_\odot \in [11.30,12.0]$ and
$\log M_\ast/h^{-2}_{70}M_\odot \in [11.40,12.0]$ (subsamples B and C
respectively). The subsamples of higher stellar mass thresholds
progressively display the higher amplitudes in the clustering and weak
lensing signals (Figures~\ref{fig:wp_fix_z} and
\ref{fig:cosmo_sys}). These results suggest that the CMASS galaxies of
higher stellar masses tend to reside in more massive halos. On the other
hand, we found the weak redshift dependence of the signals for
each subsample (Figures~\ref{fig:wp_fix_mstel} and \ref{fig:cmass_lensing_redshift_subsamples}). This result allows us to
employ an effectively signal redshift bin over the redshift range when
making the model interpretation.

In Paper II, we will use the clustering and lensing measurements of
the CMASS galaxies to explore the dark matter-galaxy connection in the
framework of the halo model. By fitting the halo occupation parameters
and cosmological parameters to both of these observables simultaneously,
we will explore the physical nature of CAMSS galaxies and their host
dark matter halos as well as constrain cosmological parameters. 


\section*{Acknowledgments}
We greatly thank Hong Guo, Anatoly Klypin, Francisco Prada, Ramin
Skibba, and Simon White for useful comments during the SDSS internal
review. HM and RM greatly thank Alexie Leauthaud, Melanie
Simet, Lance Miller, Catherine Heymans, Ludovic Van Waerbeke, and
Hendrik Hildebrandt for useful discussion about the CFHTLenS catalog.

HM is supported by Japan Society for the Promotion of Science
(JSPS) Postdoctoral Fellowships for Research Abroad and JSPS Research Fellowships for Young Scientists. MT is supported in
part by JSPS KAKENHI (Grant Number: 23340061), by World Premier
International Research Center Initiative (WPI Initiative), MEXT, Japan,
and by the FIRST program `Subaru Measurements of Images and Redshifts
(SuMIRe)', CSTP, Japan.
RM is supported in part by the Department of
Energy Early Career Award program.
This work was supported in part by the National Science Foundation under Grant No. PHYS-1066293 and the hospitality of the Aspen Center for Physics.

Funding for SDSS-III has been provided by the Alfred P. Sloan Foundation, the Participating Institutions, the National Science Foundation, and the U.S. Department of Energy Office of Science. The SDSS-III web site is http://www.sdss3.org/.

SDSS-III is managed by the Astrophysical Research Consortium for the
Participating Institutions of the SDSS-III Collaboration including the
University of Arizona, the Brazilian Participation Group, Brookhaven
National Laboratory, University of Cambridge, Carnegie Mellon
University, University of Florida, the French Participation Group, the
German Participation Group, Harvard University, the Instituto de
Astrofisica de Canarias, the Michigan State/Notre Dame/JINA
Participation Group, Johns Hopkins University, Lawrence Berkeley
National Laboratory, Max Planck Institute for Astrophysics, Max Planck
Institute for Extraterrestrial Physics, New Mexico State University, New
York University, Ohio State University, Pennsylvania State University,
University of Portsmouth, Princeton University, the Spanish
Participation Group, University of Tokyo, University of Utah, Vanderbilt
University, University of Virginia, University of Washington, and Yale
University.

This work is based on observations obtained with MegaPrime/MegaCam, a joint
project of CFHT and CEA/IRFU, at the Canada-France-Hawaii Telescope (CFHT)
which is operated by the National Research Council (NRC) of Canada, the
Institut National des Sciences de l'Univers of the Centre National de la
Recherche Scientifique (CNRS) of France, and the University of Hawaii. This
research used the facilities of the Canadian Astronomy Data Centre operated by
the National Research Council of Canada with the support of the Canadian Space
Agency. CFHTLenS data processing was made possible thanks to significant
computing support from the NSERC Research Tools and Instruments grant program.


\bibliographystyle{apj}
\bibliography{paper}

\appendix
\section{Calculation of LRG Lensing Signal}
\label{app:LRG_lensing_signal} 

In this section we describe details of our measurement of the SDSS LRG
lensing signal shown in Figure~\ref{fig:lens_signal}. We select LRGs
\citep{Eisenstein:2001} in the regions overlapping with CFHTLenS fields
from the SDSS DR7 catalog \citep{Abazajian:2009}. Out of 62,081 LRGs of
the entire SDSS fields, 534 LRGs are selected. The number of LRGs in
each CFHTLenS field is 149 in W1, 0 in W2, 338 in W3, and 47 in W4,
respectively.  We use the same weighting scheme as that in RM13,
which is given as
\begin{equation}
\label{eq:lens_weight}
w_\rml=\frac{w_{\rm rad}w_{\rm fc}}{C},
\end{equation}
where $w_{\rm rad}$ is the weight to account for the radial selection
function, $w_{\rm fc}$ is the fiber collision weight, and $C$ is the
``sector completeness'' that accounts for the redshift success rate.
These weights are available from the publicly-available LRG catalog (see
Appendix~A in \citealt{Kazin:2010} for the details).

\begin{figure*}
\begin{center}
\includegraphics[width=8cm]{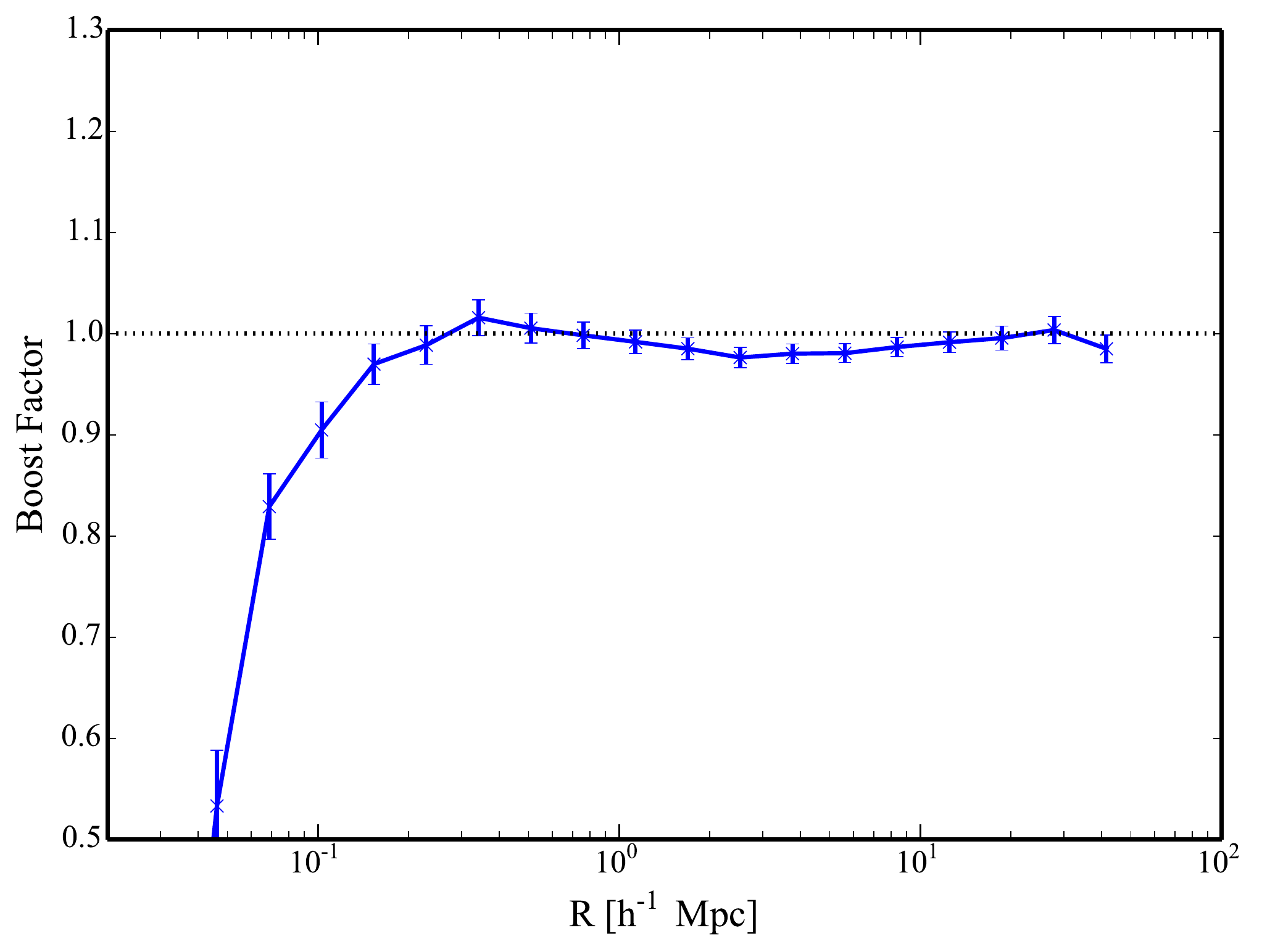}
\includegraphics[width=8cm]{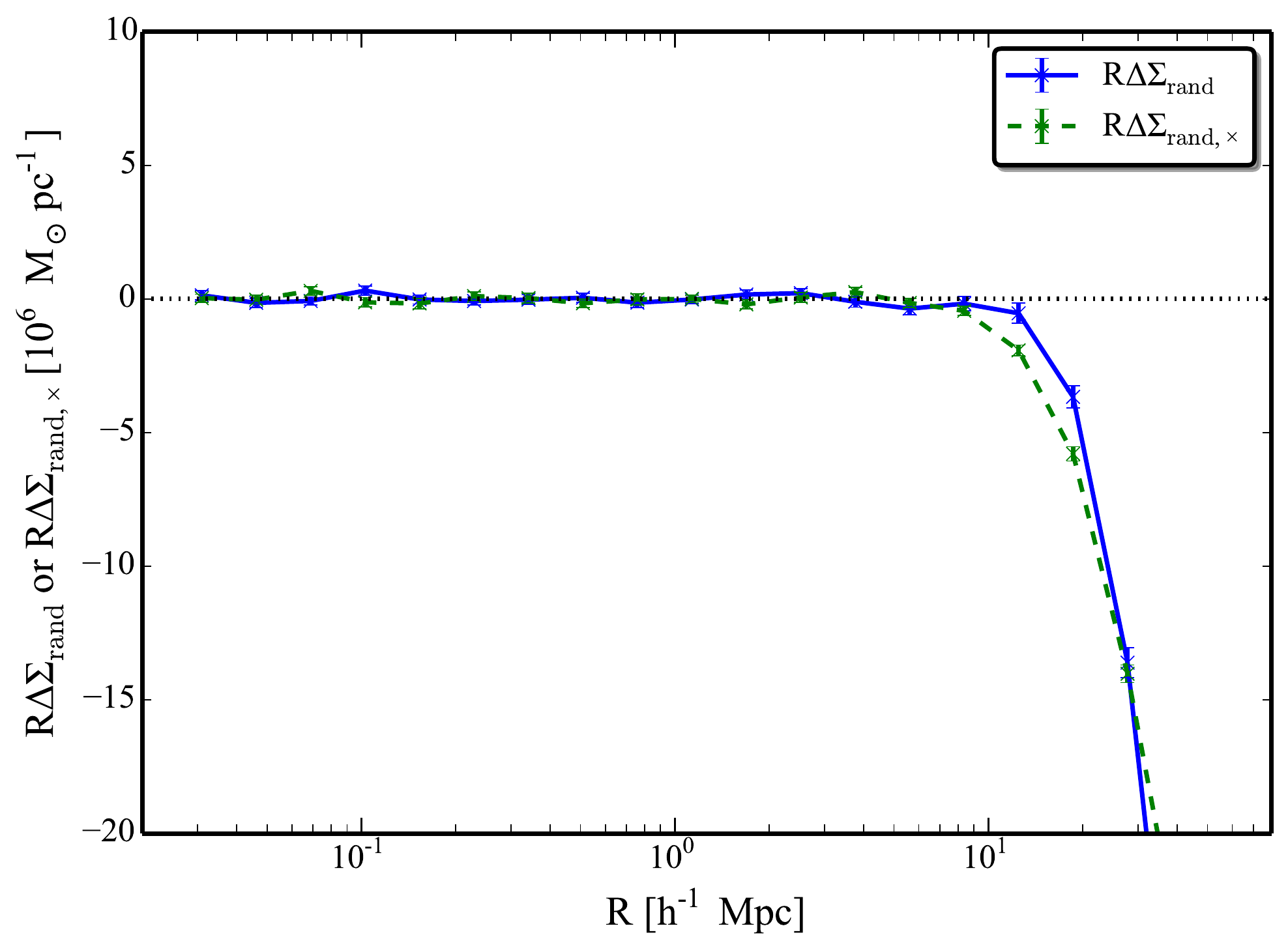} 

\caption{{\it Left panel}: A boost factor of LRGs. Although the boost factor is
 slightly smaller than unity at intermediate scales, this does not affect
 the consistency between our CFHTLenS LRG signal and the SDSS LRG lensing
 signal described in Section~\ref{subsec:CMASS_galaxy_signal} (see text
 for discussion). {\it Right:} The random signal of LRGs. As seen in
 the CMASS galaxy measurement, the random signal is not consistent with
 zero, which implies a coherent PSF anisotropy at the edge of
 the field-of-view is not fully corrected. We corrected for this effect by
 subtracting the random signal from a measured lensing signal.}
 \label{fig:systematics_LRG}
\end{center}
\end{figure*}

We perform systematic tests of the LRG lensing signal using random
catalogs as described in Section~\ref{subsec:systematic_tests}. We
generate 100 realizations of the random catalogs for the overlapping
regions of the SDSS and CFHTLenS fields.  The left panel of
Figure~\ref{fig:systematics_LRG} shows the boost factor for the LRG
sample. As we found from the boost factor for our subsample of the CMASS
galaxies in Figure~\ref{fig:systematics}, the small scales displays
$B(R)<1$. Thus we do not use the lensing measurements at $R<150 \kpch$
(up to the 4th bin).  While the boost factor is consistent with
unity at the larger radii, it is slightly smaller than unity by a few
per cent at the intermediate radii.  Since our lensing signal remains
consistent with RM13 (Figure~\ref{fig:lens_signal}) within this offset, we
do not adopt the boost factor correction for the LRG lensing signals.
The right panel shows the lensing measurements for the random catalogs
of LRGs. As seen for the measurements of the CMASS random catalogs in
Figure~\ref{fig:systematics}, the large radii show non-zero signals,
indicating residual systematics in the shape measurements (see
Section~\ref{sec:random} for the discussion).  We correct for the LRG
lensing signal by subtracting the random signal from the measured
signal. Figure~\ref{fig:lens_signal} shows the measurement of LRG lensing
signal, compared to the measurement in RM13 that was done using the
independent data, the SDSS imaging galaxies, for the galaxy shape
measurements. The CFHTLenS and SDSS measurements for LRG lensing show an
excellent agreement with each other.

\end{document}